\documentclass[a4paper,11pt]{article}
\usepackage{pos}
\usepackage{xspace}
\usepackage{mathtools}
\usepackage{mwe}
\usepackage{subcaption}
\usepackage{graphicx}

\usepackage{textcomp}
\usepackage{enumitem}
\usepackage{hyperref}

\usepackage[
backend=biber,
style=numeric,
sorting=none
]{biblatex}
\usepackage{orcidlink}

\def\pyirf{\texttt{pyirf}\xspace}
\def\hairspace{\hskip 0.1em\relax}

\title{
  Interpolation of Instrument Response Functions for the Cherenkov Telescope Array in 
  the Context of \pyirf
}
\ShortTitle{Interpolation of IRFs for CTA with \pyirf}

\manuallySeparateAuthors

\author*[a]{R.\hairspace M.~Dominik\orcidlink{0000-0003-4168-7200}\hairspace}
\author[a]{, M.~Linhoff\orcidlink{0000-0001-7993-8189}\hairspace}
\author[b]{ and J.~Sitarek\orcidlink{0000-0002-1659-5374}\hairspace}
\author{ for the CTA Consortium}

\affiliation[a]{TU Dortmund University, Department of Physics\\
  Otto-Hahn-Stra{\ss}e 4a, D-44227, Germany}

\affiliation[b]{University of Lodz, Faculty of Physics and Applied Informatics\\
Pomorska 149/153, 90-236 Lodz, Poland}

\emailAdd{rune.dominik@tu-dortmund.de}
\emailAdd{maximilian.linhoff@tu-dortmund.de}
\emailAdd{jsitarek@uni.lodz.pl}

\renewcommand{\printHeadAuthors}{R.~M.~Dominik et al.}

\abstract{
  The Cherenkov Telescope Array (CTA) will be the next generation ground-based
  very-high-energy gamma-ray observatory, constituted by tens of Imaging Atmospheric
  Cherenkov Telescopes at two sites once its construction and commissioning are finished. 
  Like its predecessors, CTA relies on Instrument Response Functions (IRFs) to relate 
  the observed and reconstructed properties to the true ones of the primary gamma-ray 
  photons. 
  IRFs are needed for the proper reconstruction of spectral and spatial information of 
  the observed sources and are thus among the data products issued to the observatory 
  users. 
  They are derived from Monte Carlo simulations, depend on observation conditions like
  the telescope pointing direction or the atmospheric transparency and can evolve with 
  time as hardware ages or is replaced. 
  Producing a complete set of IRFs from simulations for every observation taken is a 
  time-consuming task and not feasible when releasing data products on short timescales. 
  Consequently, interpolation techniques on simulated IRFs are investigated to quickly 
  estimate IRFs for specific observation conditions. 
  However, as some of the IRFs constituents are given as probability distributions, 
  specialized methods are needed. 
  This contribution summarizes and compares the feasibility of multiple approaches to 
  interpolate IRF components in the context of the \pyirf python software package and 
  IRFs simulated for the Large-Sized Telescope prototype (LST-1). 
  We will also give an overview of the current functionalities implemented in \pyirf.
}

\ConferenceLogo{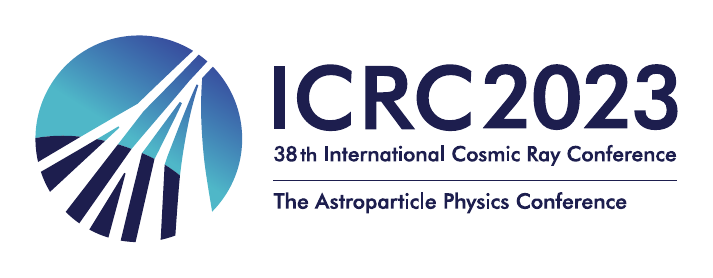}

\FullConference{%
The 38th International Cosmic Ray Conference (ICRC2023)\\
  26 July -- 3 August, 2023\\
  Nagoya, Japan}

\begin{document}
\maketitle

\section{Introduction}
Instrument Response Functions (IRFs) are crucial in analyzing data taken by Imaging Air 
Cherenkov Telescopes (IACTs) as they relate the reconstructed information and the true 
ones for incomming gamma rays. 
The first VHE gamma-ray observatory, the under construction Cherenkov Telescope Array 
Observatory (CTAO), will consist of tens of telescopes in two arrays located at the Northern
and Southern Hemisphere \cite{cta_mc_opt}. 
As CTAO is designed to detect gamma rays with energies between 20\,GeV and 300\, TeV 
with unprecedented angular and energy resolution, it has stringent requirements on 
systematic uncertainties which are dominated by how well the derived IRFs reproduce the actual 
response of the telescopes.
To manage this, extensive, time- and resource-consuming, Monte Carlo simulations will be 
needed to generate the data necessary for the computation of the final IRFs. 
In settings where time and resources are constrained, creating simulations tailored to 
each pointing position is infeasible. 
In these cases, IRF interpolation may offer a solution.
In CTAO's software ecosystem, both IRF computation and interpolation are managed by the 
\pyirf software package.

This proceeding aims to give a brief overview of the \pyirf package with a focus 
on IRF interpolation in \pyirf. 
For this, section \ref{chapter_irfs} will give a more thorough introduction to IRFs and 
section \ref{chapter_pyirf} will introduce \pyirf and its main interpolation 
functionalities.
Results of these methods on IRFs will be shown in section \ref{chapter_interpolation} 
and section \ref{chapter_conclusions} will summarize this proceeding.

\section{Instrument Response Functions}
\label{chapter_irfs}
While IACT experiments employ advanced machine learning methods to reconstruct the 
observed air showers as-good-as-possible, the result always has a finite accuracy.
Trying to estimate energy $\hat{E}$, origin as right ascension and declination 
$(\hat{\alpha}, \hat{\delta})$ and arrival time $\hat{t}$ of a gamma-ray signal,
deviations from the true values $E, \alpha$ and $\delta$ are inevitable
and that some events are not recorded or successfully reconstructed at all.
For most applications, the time measurement can be assumend to be precise so that 
$t = \hat{t}$ holds.
In its most general form, an IRF $R$ is the conditional probability relating 
the observed distribution of events $g(\hat{E}, \hat{\alpha}, \hat{\delta}, 
t)$ to the true gamma-ray signal arriving at Earth $f(E, \alpha, \delta, t)$ 
with a background $b(\hat{E}, \hat{\alpha}, \hat{\delta}, t)$ by 
\begin{equation}
  \underbrace{g(\hat{E}, \hat{\alpha}, \hat{\delta}, t)}_{\mathclap{\text{Observed 
    distribution}}} = \iiint \overbrace{%
    R(\hat{E}, \hat{\alpha}, \hat{\delta} | E, \alpha, \delta, t)
    }^{\mathclap{\text{Instrument Response}}}
    \cdot \underbrace{f(E, \alpha, \delta, t)}_{\mathclap{\text{True gamma-ray signal}}}
    \, \mathrm{d}E \, \mathrm{d}\Omega \, \mathrm{d}t
    + \overbrace{b(\hat{E}, \hat{\alpha}, \hat{\delta}, t)}^{\mathclap{
    \text{Background}}}
    \label{eq_irf}
\end{equation}
with the solid angle differential $\mathrm{d}\Omega = \sin{\delta}\,\mathrm{d}\alpha \, 
\mathrm{d}\delta$.
Even when assuming a sufficiently correct  
knowledge of $b(\hat{E}, \hat{\alpha}, \hat{\delta}, t)$, the IRF is a 
six-dimensional, time-dependent quantity $R(\hat{E}, \hat{\alpha}, \hat{\delta} | 
E, \alpha, \delta, t)$. 
As IRFs are generated from Monte Carlo simulations, where after processing both the true 
and the reconstructed values are known, it is infeasible to generate the amount of events needed 
to compute $R$ in this general form.
Since it is nevertheless needed to solve \eqref{eq_irf} and, with it, correctly reconstruct
spectral and spatial information, a dimension reduction by factorization 
\begin{equation}
        R(\hat{E}, \hat{\alpha}, \hat{\delta}| E, \alpha, \delta, t) =
        \underbrace{A_\text{eff}(E, \alpha, \delta, t)}_{\mathclap{\text{Effective Area}}}
        \cdot \overbrace{M(\hat{E} | E, \alpha, \delta, t)}^{\mathclap{
        \text{Energy Migration}}} \cdot \underbrace{\operatorname{PSF}(\hat{\alpha}, 
        \hat{\delta} | E, \alpha, \delta, t)}_{\mathclap{\text{Point Spread Function}}}
 \end{equation}
is commonly applied and IRFs are expressed as parametrizations or discretized tables of 
these components.
Thus, the 
\begin{itemize}
  \item Effective Area (\texttt{AEFF}), the combination of the experiment's sensitive 
  area and the probability of a gamma ray with some true properties to be present in 
  the data as a gamma ray after all analysis steps,
  \item Energy Migration (\texttt{EDISP}), the conditional probability to reconstruct a gamma ray of some 
  true properties with a certain energy $\hat{E}$ and
  \item Point Spread Function (\texttt{PSF}), the conditional probability to reconstruct a 
  gamma ray of some true properties at a certain origin $(\hat{\alpha}, \hat{\delta})$
\end{itemize}
constitute an IRF that is applicable to all analysis use-cases.
Further simplifications can be made for the case of a point source, where events can be 
selected around the assumed point source position using potential $\hat{E}$-dependent radii.
In this case, the full PSF is not needed and instead the effective area is reduced by the 
amount of non-selected events. 
The used radii are linked to the IRFs and must be stored along-side in so-called 
\texttt{RAD\_MAX}-tables.
All these components are usually assumed constant over some time window and further depend 
on observation conditions like telescope pointing or weather.
As the telescope performance is not constant over the whole field of view (FoV), all 
components are typically computed in more than one bin of FoV offset.

Allthough the process of simulating sufficient events to compute IRFs is time and 
resource-consuming, it assures the best possible results.
On the other hand, there are circumstances where IRFs are needed on short time scales
for next-day or even real-time analyses.
This might be the case when observing an unexpected, transient event, and a preliminary 
analysis is needed to alert other experiments for follow-up observations.
Interpolation between IRFs, precomputed for some observation conditions, is a possible 
solution in these cases and, if sufficiently performant, the less resource intensive 
solution to IRF computation.

\section{The \pyirf Package}
\label{chapter_pyirf}
The open-source software package \pyirf \cite{pyirf_zenodo} is a python library to 
compute IRFs and, derived from them, sensitivities. 
It is developed on Github\footnote{\url{https://github.com/cta-observatory/pyirf}} and 
released to PyPI\footnote{\url{https://pypi.org/project/pyirf}} and 
conda-forge\footnote{\url{https://anaconda.org/conda-forge/pyirf}}.
While the main use case will lie within CTAO's analysis framework, \pyirf does not rely on 
specialized input formats but rather \texttt{astropy}'s \texttt{QTable} 
\cite{astropy_zenodo} and is thus, by design, usable with any IACT experiment.
The internal representation of IRFs and, with this, \pyirf's output is, on the other hand, 
compatible with the \textit{data formats for gamma-ray astronomy} (GADF) 
\cite{OGADF}.
In the CTA software ecosystem, \pyirf processes the so-called DL2 stage of gamma-ray 
Monte Carlo simulations containing reconstructed air-shower events generated by 
\texttt{ctapipe} \cite{ctapipe_zenodo} to obtain IRFs. 
With the initial DL2 data and observatory metadata, IRFs are issued to the 
scientific users of CTA as DL3 data and are ready for usage with CTA's science tools
like \texttt{gammapy} \cite{gammapy_proc}.
Alongside the computation of IRFs, \pyirf also provides sensitivity and significance 
calculation and IRF interpolation, the focus of this proceeding. 



\subsection{Interpolate IRFs from Existing IRFs}
\label{chapter_pyirf_interpolation}
To interpolate IRFs and contrary to other \pyirf functionalities, the user does not 
have to supply reconstructed shower events but computed IRFs on a grid in some 
observation parameters that influence the telescope performance and, therefore, the IRFs.
One such choice might be a grid in zenith angular distance and the angle between the 
telescope pointing direction and the geomagnetic field.
Both parameters influence shower development in the atmosphere.
Currently, \pyirf supports the interpolation of \texttt{AEFF}, \texttt{EDISP},
\texttt{PSF} and \texttt{RAD\_MAX} quantities according to the GADF definitions 
and with that both full enclosure and point like IRFs. 

While \texttt{AEFF} and \texttt{RAD\_MAX} are simple, unconnected quantities, i.e., 
each bin holds a value that is not correlated to its neighboring bins, simple, e.g. linear,
interpolation can be applied.
Contrary, \texttt{EDISP} and \texttt{PSF} represent discretized probability density functions; therefore,
both need specialized methods to maintain their internal consistency.
Such methods have already been employed in high-energy particle physics, \pyirf offers 
two, Quantile Interpolation \cite{QuantileInterpRead, QuantileInterpHollister} and 
Moment Morphing \cite{MomentMorph}.

In short summary, Quantile Interpolation utilizes that there exist points $x_{i}$ where 
the template distribution's cumulative distribution functions (CDF) $F_{i}$ give the same
value $F_{i}(x_{i}) = y$. 
The target distribution's CDF is then constructed to also return this value at a linear 
interpolation of the $x_{i}$. 
However, to obtain this position, the template distribution's quantile functions are needed 
and ultimately interpolated.
The second method, Moment Morphing, is based on a Taylor expansion on the parameter grid 
used to find suitable interpolation coefficients. 
These, in return, are used to construct a linear combination of the template distributions; 
the interpolated result. 
To do so correctly, it is necessary to account for the template distribution's varying mean and 
standard deviation by transforming them to common values first.
Both methods need to be adapted to be usable with discretized distributions.
For Quantile Interpolation, this includes the usage of empirical distribution functions
and linear interpolation to obtain an estimate for the quantile functions. 
Moment Morphing needed approximating versions of mean and standard deviation computation 
and a look-up method to evaluate the template distributions at the transformed 
values.

While \pyirf's Quantile Interpolation can, in principle, be applied to arbitrary 
parameter grid dimensions if the function interpolating the quantiles is chosen 
appropriately, Moment Morphing is currently limited to one or two-dimensional
grids as the computation of the interpolation coefficients is dimension-dependent.
On the other hand, our adapted version of Moment Morphing offers the possibility to 
extrapolate beyond the parameter grid convex hull. 
For this, two methods are implemented for extrapolation: Nearest Simplex and Visible  
Edges Extrapolation.
The first one extrapolates from the nearest triangular simplex as seen from a point 
outside the parameter grid, resulting in a non-continuous extrapolation function. 
This is solved by the second method, which computes extrapolations from all visible 
simplices, computing coefficients according to visible edges blending as discussed in 
\cite{VisbileEdge} and then again using these coefficients in the Moment Morphing 
procedure to compute the actual estimation by combining all extrapolations.
While a continuous extrapolation is desirable, we provide both methods to leave 
it to the user's discretion to utilize the additional assumptions needed for visible 
edges blending. 
For one-dimensional parameter grids, both methods are equivalent.
The same holds for points outside two-dimensional grids where only one edge and thus 
triangular grid simplex is visible.
In all cases, we advise caution, as our extrapolation implementations are extensions 
of the Moment Morphing procedure that utilize negative interpolation coefficients.
It is easily possible for bin entries to become negative and thus ill-defined when 
overlaying the template distributions with negative coefficients. 
While this effect is minor and, to some extent, accountable by cutting off affected bins 
for small extrapolation distances, high extrapolation distances result in meaningless 
estimations.
Extrapolation should thus be avoided by extending the template distribution grid to 
include all desired target points.
For convenience, we also provide dummy extrapolation using the nearest neighbor approach.  

\begin{figure}
    \centering
    \begin{subfigure}{0.495\textwidth}
        \includegraphics{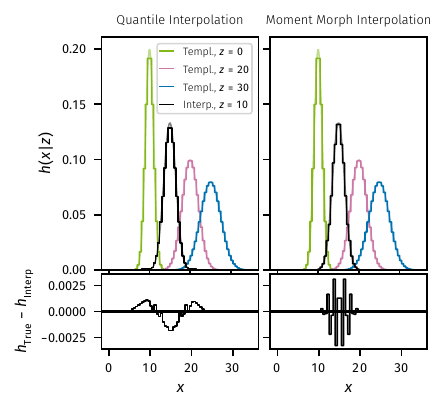}
    \end{subfigure}
    \hfill
    \begin{subfigure}{0.495\textwidth}
        \includegraphics{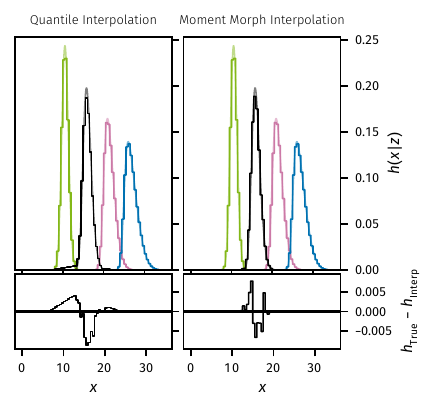}
    \end{subfigure}
    \caption{Interpolation algorithms applied to Gaussian (\textbf{left}) and skewed
        Gaussian (\textbf{right}) distributions whose parameters vary linearly with $z$.
        Three template histograms were created from the distribution's CDFs and used to 
        interpolate to the black distribution. 
        True distributions are given as solid lines.}
    \label{figure_1Dgaussians}
\end{figure}

\begin{figure}
    \centering
    \begin{subfigure}{0.495\textwidth}
        \includegraphics{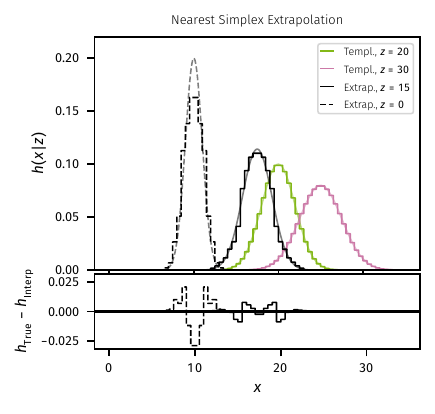}
    \end{subfigure}
    \hfill
    \begin{subfigure}{0.495\textwidth}
        \includegraphics{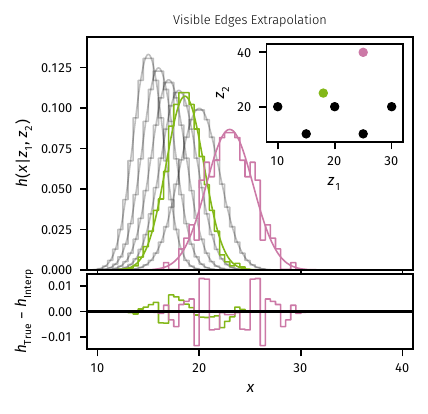}
    \end{subfigure}
    \caption{Extrapolation applied to Gaussian on a one- (\textbf{left}) and 
        two-dimensional grid (\textbf{right}).
        Gaussian distributions whose parameters vary linearly with 
        $z$ and $z_{1}$ and $z_{2}$, respectively.
        Two and five template histograms were created from the distribution's CDFs 
        and used to extrapolate to the black distribution.}
    \label{figure_1Dgaussians_extrapolation}
\end{figure}

Quantile Interpolation and Moment Morphing have been applied to one-dimensional 
Gaussians and skewed Gaussians in Fig.\ref{figure_1Dgaussians}, performing well 
in this simple demonstration. 
However, skewed distributions seem harder to interpolate, which is easily 
explainable for Moment Morphing as it only accounts for first and second-order 
moments by design. 
The good performance holds for comparably small extrapolation 
distances, as shown in Fig.\ref{figure_1Dgaussians_extrapolation}.
As expected, the result worsens with increased distance.

\section{IRF Interpolation}
\label{chapter_interpolation}
\begin{figure}
    \centering
    \begin{subfigure}{0.495\textwidth}
        \includegraphics{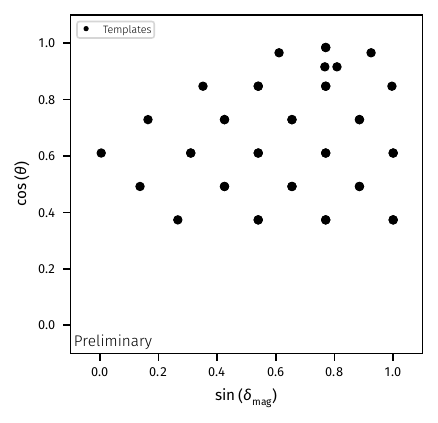}
    \end{subfigure}
    \hfill
    \begin{subfigure}{0.495\textwidth}
        \includegraphics{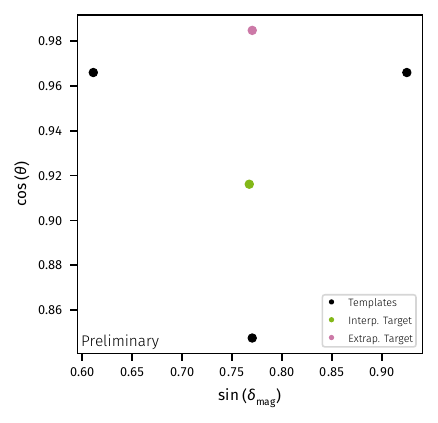}
    \end{subfigure}
    \caption{Full grid (\textbf{left}) of IRFs produced for LST-1 and 
    the subset (\textbf{right}) selected to showcase both interpolation and 
    extrapolation results.}
    \label{figure_grid}
\end{figure}

In addition to the benchmark presented in the previous section, performance
measures on actual IRFs were performed.
To do so, we used a subset of an IRF grid produced for the Large-Sized Telescope 
prototype LST-1 \cite{lst_performance}.
As mentioned in section \ref{chapter_pyirf_interpolation}, the grid is produced
in zenith distance $\theta$ and the angle between telescope pointing and the geomagnetic
field $\delta_{\mathrm{mag}}$.
This angle is not to be confused with the astronomical declination introduced in section 
\ref{chapter_irfs}, although they share a common symbol.
To better reflect the physical development of an air shower, we use $\cos{\theta}$ 
(dependency of the atmospheres density profile and thus Cherenkov light absorption) 
and $\sin{\delta_{\mathrm{mag}}}$ (measure of the geomagnetic field effect on the shower development).
Other choices, especially for transformations of $\theta$, are possible, e.g., 
$(\cos{\theta})^{-1}$, to reflect the atmospheric depth along the shower's line of sight.
The full grid and a selected part used for the following showcase are displayed 
in Fig. \ref{figure_grid}.
Multiple grid nodes exist in this representation, where more than one Monte Carlo 
production has been made for different azimuth pointings.
In these cases, the nodes closest to the target's azimuth pointing have been selected as 
interpolation templates.

\begin{figure}
    \centering
    \begin{subfigure}{0.495\textwidth}
        \includegraphics{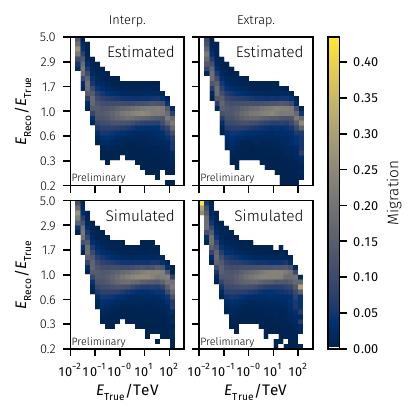}
    \end{subfigure}
    \hfill
    \begin{subfigure}{0.495\textwidth}
        \includegraphics{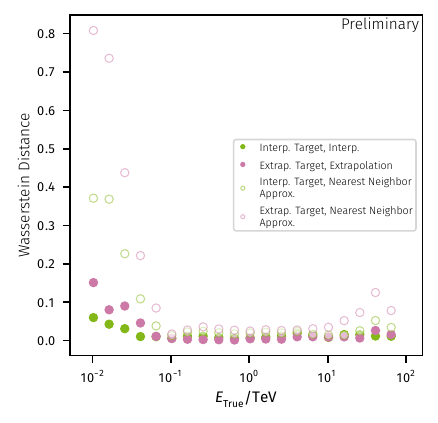}
    \end{subfigure}
    \caption{Inter- and Extrapolated energy dispersions (\textbf{left}) with 
    corresponding Wasserstein distances (\textbf{right}). 
    A simple next-neighbor interpolation was added for reference.}
    \label{figure_edisps}
\end{figure}

Testing the interpolation on the selected subset of the full grid, we find good agreement
between simulated and estimated energy dispersions, as shown in Fig. \ref{figure_edisps}. 
The estimated results outperform a nearest-neighbor approach, especially for low- and high 
energies, and extrapolation, in this case, is feasible. 
It can be seen in the extrapolation case of Fig. \ref{figure_edisps} that the lowest 
true energy bin is missing in the estimation. 
This occurs when one of the used templates has, due to a higher energy threshold, 
no corresponding IRF values drived in the respective bin.
This is caused by an increasing zenith angle as the amount of atmosphere 
traversed by a gamma ray grows and thus the possibility for the primary particle 
to interact before reaching the telescope increases. 
Primary particles of these energies are thus less and less often detected until these energies 
are lost. 
Consequently, this problem is more frequent when high interpolation distances along the 
zenith angle are chosen.
A densely populated grid thus minimizes the effect, although there will always be 
translations in the grid's parameter space where one bin becomes empty.

\begin{figure}
    \centering
    \begin{subfigure}{0.495\textwidth}
        \includegraphics{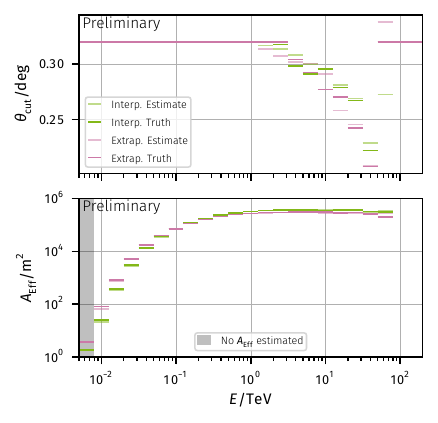}
    \end{subfigure}
    \hfill
    \begin{subfigure}{0.495\textwidth}
        \includegraphics{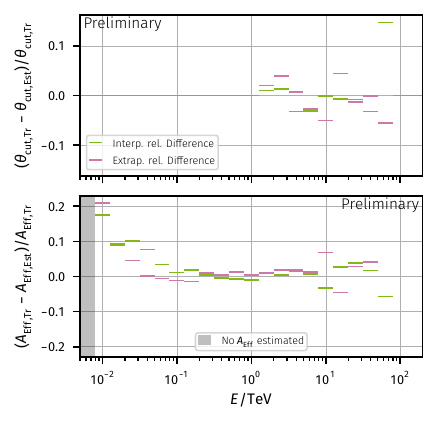}
    \end{subfigure}
    \caption{Inter- and Extrapolated \texttt{RAD\_MAX} and \texttt{AEFF} tables (\textbf{left}) 
    with relative error (\textbf{right}).}
    \label{figure_aeff_rad_max}
\end{figure}

As for the estimation of AEFF and RAD\_MAX tables (see Fig. 
\ref{figure_aeff_rad_max}), we find that especially effective areas are estimated with 
minor errors compared to the actual values.
As with the energy dispersions, the lowest energy bin was missing in some templates and 
could not be computed.
RAD\_MAX tables, on the other hand, with oszillating errors, perform more erratic than 
other components. 
We account this to the nature of producing these values, optimizing a cut-value instead 
of comparing simulated and reconstructed quantities. 
RAD\_MAX tables thus violate the assumption of being dependent on the chosen set of 
grid parameters. 
It is, however, to be checked, if these interpolation results in a meaningful IRF at all
as it is not guaranteed that AEFF and RAD\_MAX values are matching afterwards.

\section{Conclusion}
\label{chapter_conclusions}
In this proceeding, we have given a brief overview of \pyirf and explicitly
discussed the therein-provided IRF inter- and extrapolation methods.
We have shown IRF inter- and extrapolation to be usable with toy data sets and actual 
IRFs.
Especially for quantities independent from any user optimization, like effective area
and energy dispersion, the generated IRFs are reasonable estimations of the truth and 
outperform simple, next-neighbor approaches.
Errors are minor but present, especially in low- or high-energy bins, and 
increase with inter- or extrapolation distance.
Edge effects may occur if the interpolation templates partially miss some 
energy ranges, e.g., low true energy entries missing in energy dispersions for high 
zenith angles.
User optimized quantities, like \texttt{RAD\_MAX}-tables, are less suited for 
interpolation.
In general, however, IRF interpolation has the potential to produce 
sufficiently accurate results for short time scale analyses.


\section*{Acknowledgments}
\scriptsize{
\vspace{-0.75\baselineskip}
This work was conducted in the context of the CTA Consortium and CTA Observatory.
We gratefully acknowledge financial support from the agencies and organizations listed here:
\url{https://www.cta-observatory.org/consortium_acknowledgments/}}
\vspace{-0.5\baselineskip}
\AtNextBibliography{\scriptsize}
\scriptsize{
\vspace{-0.75\baselineskip}
}
\printbibliography
%
%
%

\pagestyle{empty}
\centering\LARGE
The CTA Consortium\\[0.5cm]
\normalsize
\raggedright
  \mbox{K.~Abe$^{\ref{AFFIL::JapanUTokai}}$}, 
  \mbox{S.~Abe$^{\ref{AFFIL::JapanUTokyoICRR}}$\orcidlink{0000-0001-7250-3596}}, 
  \mbox{A.~Acharyya$^{\ref{AFFIL::USAUAlabamaTuscaloosa}}$\orcidlink{0000-0002-2028-9230}}, 
  \mbox{R.~Adam$^{\ref{AFFIL::FranceOCotedAzur},\ref{AFFIL::FranceLLREcolePolytechnique}}$}, 
  \mbox{A.~Aguasca-Cabot$^{\ref{AFFIL::SpainICCUB}}$\orcidlink{0000-0001-8816-4920}}, 
  \mbox{I.~Agudo$^{\ref{AFFIL::SpainIAACSIC}}$\orcidlink{0000-0002-3777-6182}}, 
  \mbox{J.~Alfaro$^{\ref{AFFIL::ChileUPontificiaCatolicadeChile}}$}, 
  \mbox{N.~Alvarez-Crespo$^{\ref{AFFIL::SpainUCMAltasEnergias}}$}, 
  \mbox{R.~Alves~Batista$^{\ref{AFFIL::SpainIFTUAMCSIC}}$\orcidlink{0000-0003-2656-064X}}, 
  \mbox{J.-P.~Amans$^{\ref{AFFIL::FranceObservatoiredeParis}}$}, 
  \mbox{E.~Amato$^{\ref{AFFIL::ItalyOArcetri}}$\orcidlink{0000-0002-9881-8112}}, 
  \mbox{F.~Ambrosino$^{\ref{AFFIL::ItalyORoma}}$}, 
  \mbox{E.~O.~Ang\"uner$^{\ref{AFFIL::TurkeyTubitak}}$\orcidlink{0000-0002-4712-4292}}, 
  \mbox{L.~A.~Antonelli$^{\ref{AFFIL::ItalyORoma}}$\orcidlink{0000-0002-5037-9034}}, 
  \mbox{C.~Aramo$^{\ref{AFFIL::ItalyINFNNapoli}}$}, 
  \mbox{C.~Arcaro$^{\ref{AFFIL::ItalyINFNPadova}}$\orcidlink{0000-0002-1998-9707}}, 
  \mbox{L.~Arrabito$^{\ref{AFFIL::FranceLUPMUMontpellier}}$\orcidlink{0000-0003-4727-7288}}, 
  \mbox{K.~Asano$^{\ref{AFFIL::JapanUTokyoICRR}}$}, 
  \mbox{J.~Aschersleben$^{\ref{AFFIL::NetherlandsUGroningen}}$\orcidlink{0000-0002-6097-7898}}, 
  \mbox{H.~Ashkar$^{\ref{AFFIL::FranceLLREcolePolytechnique}}$\orcidlink{0000-0002-2153-1818}}, 
  \mbox{L.~Augusto~Stuani$^{\ref{AFFIL::BrazilIFSCUSaoPaulo}}$}, 
  \mbox{D.~Baack$^{\ref{AFFIL::GermanyUDortmundTU}}$\orcidlink{0000-0002-2311-4460}}, 
  \mbox{M.~Backes$^{\ref{AFFIL::NamibiaUNamibia},\ref{AFFIL::SouthAfricaNWU}}$\orcidlink{0000-0002-9326-6400}}, 
  \mbox{C.~Balazs$^{\ref{AFFIL::AustraliaUMonash}}$\orcidlink{0000-0001-7154-1726}}, 
  \mbox{M.~Balbo$^{\ref{AFFIL::SwitzerlandUGenevaISDC}}$\orcidlink{0000-0002-6556-3344}}, 
  \mbox{A.~Baquero~Larriva$^{\ref{AFFIL::SpainUCMAltasEnergias},\ref{AFFIL::EcuadorUAzuay}}$\orcidlink{0000-0002-1757-5826}}, 
  \mbox{V.~Barbosa~Martins$^{\ref{AFFIL::GermanyDESY}}$\orcidlink{0000-0002-5085-8828}}, 
  \mbox{U.~Barres~de~Almeida$^{\ref{AFFIL::BrazilCBPF},\ref{AFFIL::BrazilIAGUSaoPaulo}}$\orcidlink{0000-0001-7909-588X}}, 
  \mbox{J.~A.~Barrio$^{\ref{AFFIL::SpainUCMAltasEnergias}}$\orcidlink{0000-0002-0965-0259}}, 
  \mbox{D.~Bastieri$^{\ref{AFFIL::ItalyUPadovaandINFN}}$\orcidlink{0000-0002-6954-8862}}, 
  \mbox{P.~I.~Batista$^{\ref{AFFIL::GermanyDESY}}$\orcidlink{0000-0001-8138-1391}}, 
  \mbox{I.~Batkovic$^{\ref{AFFIL::ItalyUPadovaandINFN}}$\orcidlink{0000-0002-1209-2542}}, 
  \mbox{R.~Batzofin$^{\ref{AFFIL::GermanyUPotsdam}}$\orcidlink{0000-0002-5797-3386}}, 
  \mbox{J.~Baxter$^{\ref{AFFIL::JapanUTokyoICRR}}$\orcidlink{0009-0004-9545-794X}}, 
  \mbox{G.~Beck$^{\ref{AFFIL::SouthAfricaUWitwatersrand}}$\orcidlink{0000-0003-4916-4914}}, 
  \mbox{J.~Becker~Tjus$^{\ref{AFFIL::GermanyUBochum}}$\orcidlink{0000-0002-1748-7367}}, 
  \mbox{L.~Beiske$^{\ref{AFFIL::GermanyUDortmundTU}}$\orcidlink{0009-0004-7097-0122}}, 
  \mbox{D.~Belardinelli$^{\ref{AFFIL::ItalyINFNRomaTorVergata}}$\orcidlink{0000-0001-9332-5733}}, 
  \mbox{W.~Benbow$^{\ref{AFFIL::USACfAHarvardSmithsonian}}$\orcidlink{0000-0003-2098-170X}}, 
  \mbox{E.~Bernardini$^{\ref{AFFIL::ItalyUPadovaandINFN}}$\orcidlink{0000-0003-3108-1141}}, 
  \mbox{J.~Bernete~Medrano$^{\ref{AFFIL::SpainCIEMAT}}$\orcidlink{0000-0002-8108-7552}}, 
  \mbox{K.~Bernl\"ohr$^{\ref{AFFIL::GermanyMPIK}}$\orcidlink{0000-0001-8065-3252}}, 
  \mbox{A.~Berti$^{\ref{AFFIL::GermanyMPP}}$\orcidlink{0000-0003-0396-4190}}, 
  \mbox{V.~Beshley$^{\ref{AFFIL::UkraineIAPMMLviv}}$}, 
  \mbox{P.~Bhattacharjee$^{\ref{AFFIL::FranceLAPPUSavoieMontBlanc}}$\orcidlink{0000-0002-0258-3831}}, 
  \mbox{S.~Bhattacharyya$^{\ref{AFFIL::SloveniaUNovaGoricaCAC}}$\orcidlink{0000-0002-6569-5953}}, 
  \mbox{B.~Bi$^{\ref{AFFIL::GermanyIAAT}}$\orcidlink{0000-0003-0455-4038}}, 
  \mbox{N.~Biederbeck$^{\ref{AFFIL::GermanyUDortmundTU}}$\orcidlink{0000-0003-3708-9785}}, 
  \mbox{A.~Biland$^{\ref{AFFIL::SwitzerlandETHZurich}}$}, 
  \mbox{E.~Bissaldi$^{\ref{AFFIL::ItalyPolitecnicoBari},\ref{AFFIL::ItalyINFNBari}}$\orcidlink{0000-0001-9935-8106}}, 
  \mbox{O.~Blanch$^{\ref{AFFIL::SpainIFAEBIST}}$\orcidlink{0000-0002-8380-1633}}, 
  \mbox{J.~Blazek$^{\ref{AFFIL::CzechRepublicFZU}}$\orcidlink{0000-0002-5870-8947}}, 
  \mbox{C.~Boisson$^{\ref{AFFIL::FranceObservatoiredeParis}}$\orcidlink{0000-0001-5893-1797}}, 
  \mbox{J.~Bolmont$^{\ref{AFFIL::FranceLPNHEUSorbonne}}$\orcidlink{0000-0003-4739-8389}}, 
  \mbox{G.~Bonnoli$^{\ref{AFFIL::ItalyOBrera},\ref{AFFIL::ItalyINFNPisa}}$\orcidlink{0000-0003-2464-9077}}, 
  \mbox{P.~Bordas$^{\ref{AFFIL::SpainICCUB}}$}, 
  \mbox{Z.~Bosnjak$^{\ref{AFFIL::CroatiaUZagreb}}$\orcidlink{0000-0001-6536-0320}}, 
  \mbox{F.~Bradascio$^{\ref{AFFIL::FranceCEAIRFUDPhP}}$\orcidlink{0000-0002-7750-5256}}, 
  \mbox{C.~Braiding$^{\ref{AFFIL::AustraliaUAdelaide}}$}, 
  \mbox{E.~Bronzini$^{\ref{AFFIL::ItalyOASBologna}}$\orcidlink{0000-0001-8378-4303}}, 
  \mbox{R.~Brose$^{\ref{AFFIL::IrelandDIAS}}$}, 
  \mbox{A.~M.~Brown$^{\ref{AFFIL::UnitedKingdomUDurham}}$\orcidlink{0000-0003-0259-3148}}, 
  \mbox{F.~Brun$^{\ref{AFFIL::FranceCEAIRFUDPhP}}$}, 
  \mbox{G.~Brunelli$^{\ref{AFFIL::ItalyOASBologna},\ref{AFFIL::SpainIAACSIC}}$}, 
  \mbox{A.~Bulgarelli$^{\ref{AFFIL::ItalyOASBologna}}$}, 
  \mbox{I.~Burelli$^{\ref{AFFIL::ItalyUUdineandINFNTrieste}}$}, 
  \mbox{L.~Burmistrov$^{\ref{AFFIL::SwitzerlandUGenevaDPNC}}$}, 
  \mbox{M.~Burton$^{\ref{AFFIL::UnitedKingdomArmaghObservatoryandPlanetarium},\ref{AFFIL::AustraliaUNewSouthWales}}$\orcidlink{0000-0001-7289-1998}}, 
  \mbox{T.~Bylund$^{\ref{AFFIL::FranceCEAIRFUDAp}}$\orcidlink{0000-0003-2946-1313}}, 
  \mbox{P.~G.~Calisse$^{\ref{AFFIL::GermanyCTAOHeidelberg}}$}, 
  \mbox{A.~Campoy-Ordaz$^{\ref{AFFIL::SpainUABandCERESIEEC}}$\orcidlink{0000-0001-9352-8936}}, 
  \mbox{B.~K.~Cantlay$^{\ref{AFFIL::ThailandUKasetsart},\ref{AFFIL::ThailandNARIT}}$\orcidlink{0009-0002-8750-6401}}, 
  \mbox{M.~Capalbi$^{\ref{AFFIL::ItalyIASFPalermo}}$\orcidlink{0000-0002-9558-2394}}, 
  \mbox{A.~Caproni$^{\ref{AFFIL::BrazilUCruzeirodoSul}}$\orcidlink{0000-0001-9707-3895}}, 
  \mbox{R.~Capuzzo-Dolcetta$^{\ref{AFFIL::ItalyORoma}}$\orcidlink{0000-0002-6871-9519}}, 
  \mbox{C.~Carlile$^{\ref{AFFIL::SwedenLundObservatory}}$}, 
  \mbox{S.~Caroff$^{\ref{AFFIL::FranceLAPPUSavoieMontBlanc}}$\orcidlink{0000-0002-1103-130X}}, 
  \mbox{A.~Carosi$^{\ref{AFFIL::ItalyORoma}}$}, 
  \mbox{R.~Carosi$^{\ref{AFFIL::ItalyINFNPisa}}$}, 
  \mbox{M.-S.~Carrasco$^{\ref{AFFIL::FranceCPPMUAixMarseille}}$}, 
  \mbox{E.~Cascone$^{\ref{AFFIL::ItalyOCapodimonte}}$\orcidlink{0000-0002-7425-7517}}, 
  \mbox{F.~Cassol$^{\ref{AFFIL::FranceCPPMUAixMarseille}}$\orcidlink{0000-0002-0372-1992}}, 
  \mbox{N.~Castrejon$^{\ref{AFFIL::SpainUAlcala}}$}, 
  \mbox{F.~Catalani$^{\ref{AFFIL::BrazilEELUSaoPaulo}}$}, 
  \mbox{D.~Cerasole$^{\ref{AFFIL::ItalyUandINFNBari}}$\orcidlink{0000-0003-2033-756X}}, 
  \mbox{M.~Cerruti$^{\ref{AFFIL::FranceAPCUParisCite}}$\orcidlink{0000-0001-7891-699X}}, 
  \mbox{S.~Chaty$^{\ref{AFFIL::FranceAPCUParisCite}}$\orcidlink{0000-0002-5769-8601}}, 
  \mbox{A.~W.~Chen$^{\ref{AFFIL::SouthAfricaUWitwatersrand}}$\orcidlink{0000-0001-6425-5692}}, 
  \mbox{M.~Chernyakova$^{\ref{AFFIL::IrelandDCU}}$\orcidlink{0000-0002-9735-3608}}, 
  \mbox{A.~Chiavassa$^{\ref{AFFIL::ItalyINFNTorino},\ref{AFFIL::ItalyUTorino}}$\orcidlink{0000-0001-6183-2589}}, 
  \mbox{J.~Chudoba$^{\ref{AFFIL::CzechRepublicFZU}}$\orcidlink{0000-0002-6425-2579}}, 
  \mbox{C.~H.~Coimbra~Araujo$^{\ref{AFFIL::BrazilUFPR}}$}, 
  \mbox{V.~Conforti$^{\ref{AFFIL::ItalyOASBologna}}$\orcidlink{0000-0002-0007-3520}}, 
  \mbox{F.~Conte$^{\ref{AFFIL::GermanyMPIK}}$\orcidlink{0000-0002-3083-8539}}, 
  \mbox{J.~L.~Contreras$^{\ref{AFFIL::SpainUCMAltasEnergias}}$\orcidlink{0000-0001-7282-2394}}, 
  \mbox{C.~Cossou$^{\ref{AFFIL::FranceCEAIRFUDAp}}$\orcidlink{0000-0001-5350-4796}}, 
  \mbox{A.~Costa$^{\ref{AFFIL::ItalyOCatania}}$\orcidlink{0000-0003-0344-8911}}, 
  \mbox{H.~Costantini$^{\ref{AFFIL::FranceCPPMUAixMarseille}}$}, 
  \mbox{P.~Cristofari$^{\ref{AFFIL::FranceObservatoiredeParis}}$}, 
  \mbox{O.~Cuevas$^{\ref{AFFIL::ChileUdeValparaiso}}$}, 
  \mbox{Z.~Curtis-Ginsberg$^{\ref{AFFIL::USAUWisconsin}}$\orcidlink{0000-0002-0194-7576}}, 
  \mbox{G.~D'Amico$^{\ref{AFFIL::NorwayUBergen}}$}, 
  \mbox{F.~D'Ammando$^{\ref{AFFIL::ItalyRadioastronomiaINAF}}$\orcidlink{0000-0001-7618-7527}}, 
  \mbox{M.~Dadina$^{\ref{AFFIL::ItalyOASBologna}}$\orcidlink{0000-0002-7858-7564}}, 
  \mbox{M.~Dalchenko$^{\ref{AFFIL::SwitzerlandUGenevaDPNC}}$\orcidlink{0000-0002-0137-136X}}, 
  \mbox{L.~David$^{\ref{AFFIL::GermanyDESY}}$\orcidlink{0000-0003-2341-9261}}, 
  \mbox{I.~D.~Davids$^{\ref{AFFIL::NamibiaUNamibia}}$\orcidlink{0000-0002-6476-964X}}, 
  \mbox{F.~Dazzi$^{\ref{AFFIL::ItalyINAF}}$\orcidlink{0000-0001-5409-6544}}, 
  \mbox{A.~De~Angelis$^{\ref{AFFIL::ItalyUPadovaandINFN}}$}, 
  \mbox{M.~de~Bony~de~Lavergne$^{\ref{AFFIL::FranceCEAIRFUDAp}}$\orcidlink{0000-0002-4650-1666}}, 
  \mbox{V.~De~Caprio$^{\ref{AFFIL::ItalyOCapodimonte}}$\orcidlink{0000-0002-4587-8963}}, 
  \mbox{G.~De~Cesare$^{\ref{AFFIL::ItalyOASBologna}}$\orcidlink{0000-0003-0869-7183}}, 
  \mbox{E.~M.~de~Gouveia~Dal~Pino$^{\ref{AFFIL::BrazilIAGUSaoPaulo}}$\orcidlink{0000-0001-8058-4752}}, 
  \mbox{B.~De~Lotto$^{\ref{AFFIL::ItalyUUdineandINFNTrieste}}$\orcidlink{0000-0003-3624-4480}}, 
  \mbox{M.~De~Lucia$^{\ref{AFFIL::ItalyINFNNapoli}}$\orcidlink{0000-0002-0519-9149}}, 
  \mbox{R.~de~Menezes$^{\ref{AFFIL::ItalyINFNTorino},\ref{AFFIL::ItalyUTorino}}$\orcidlink{0000-0001-5489-4925}}, 
  \mbox{M.~de~Naurois$^{\ref{AFFIL::FranceLLREcolePolytechnique}}$\orcidlink{0000-0002-7245-201X}}, 
  \mbox{E.~de~Ona~Wilhelmi$^{\ref{AFFIL::GermanyDESY}}$\orcidlink{0000-0002-5401-0744}}, 
  \mbox{N.~De~Simone$^{\ref{AFFIL::GermanyDESY}}$}, 
  \mbox{V.~de~Souza$^{\ref{AFFIL::BrazilIFSCUSaoPaulo}}$}, 
  \mbox{L.~del~Peral$^{\ref{AFFIL::SpainUAlcala}}$}, 
  \mbox{M.~V.~del~Valle$^{\ref{AFFIL::BrazilIAGUSaoPaulo}}$\orcidlink{0000-0002-5444-0795}}, 
  \mbox{E.~Delagnes$^{\ref{AFFIL::FranceCEAIRFUDEDIP}}$}, 
  \mbox{A.~G.~Delgado~Giler$^{\ref{AFFIL::BrazilIFSCUSaoPaulo},\ref{AFFIL::NetherlandsUGroningen}}$\orcidlink{0000-0003-2190-9857}}, 
  \mbox{C.~Delgado$^{\ref{AFFIL::SpainCIEMAT}}$\orcidlink{0000-0002-7014-4101}}, 
  \mbox{M.~Dell'aiera$^{\ref{AFFIL::FranceLAPPUSavoieMontBlanc}}$\orcidlink{0000-0002-5221-0240}}, 
  \mbox{R.~Della~Ceca$^{\ref{AFFIL::ItalyOBrera}}$\orcidlink{0000-0001-7551-2252}}, 
  \mbox{M.~Della~Valle$^{\ref{AFFIL::ItalyOCapodimonte}}$}, 
  \mbox{D.~della~Volpe$^{\ref{AFFIL::SwitzerlandUGenevaDPNC}}$\orcidlink{0000-0001-8530-7447}}, 
  \mbox{D.~Depaoli$^{\ref{AFFIL::GermanyMPIK}}$\orcidlink{0000-0002-2672-4141}}, 
  \mbox{A.~Dettlaff$^{\ref{AFFIL::GermanyMPP}}$}, 
  \mbox{T.~Di~Girolamo$^{\ref{AFFIL::ItalyUNapoli},\ref{AFFIL::ItalyINFNNapoli}}$\orcidlink{0000-0003-2339-4471}}, 
  \mbox{A.~Di~Piano$^{\ref{AFFIL::ItalyOASBologna}}$}, 
  \mbox{F.~Di~Pierro$^{\ref{AFFIL::ItalyINFNTorino}}$\orcidlink{0000-0003-4861-432X}}, 
  \mbox{R.~Di~Tria$^{\ref{AFFIL::ItalyUandINFNBari}}$\orcidlink{0009-0007-1088-5307}}, 
  \mbox{L.~Di~Venere$^{\ref{AFFIL::ItalyINFNBari}}$}, 
  \mbox{C.~D{\'\i}az-Bahamondes$^{\ref{AFFIL::ChileUPontificiaCatolicadeChile}}$}, 
  \mbox{C.~Dib$^{\ref{AFFIL::ChileUTecnicaFedericoSantaMaria}}$\orcidlink{0000-0003-4146-906X}}, 
  \mbox{S.~Diebold$^{\ref{AFFIL::GermanyIAAT}}$\orcidlink{0000-0002-8042-2443}}, 
  \mbox{R.~Dima$^{\ref{AFFIL::ItalyUPadovaandINFN}}$}, 
  \mbox{A.~Dinesh$^{\ref{AFFIL::SpainUCMAltasEnergias}}$}, 
  \mbox{A.~Djannati-Ata{\"\i}$^{\ref{AFFIL::FranceAPCUParisCite}}$\orcidlink{0000-0002-4924-1708}}, 
  \mbox{J.~Djuvsland$^{\ref{AFFIL::NorwayUBergen}}$\orcidlink{0000-0002-6488-8219}}, 
  \mbox{A.~Dom{\'\i}nguez$^{\ref{AFFIL::SpainUCMAltasEnergias}}$}, 
  \mbox{R.~M.~Dominik$^{\ref{AFFIL::GermanyUDortmundTU}}$\orcidlink{0000-0003-4168-7200}}, 
  \mbox{A.~Donini$^{\ref{AFFIL::ItalyORoma}}$\orcidlink{0000-0002-3066-724X}}, 
  \mbox{D.~Dorner$^{\ref{AFFIL::GermanyUWurzburg},\ref{AFFIL::SwitzerlandETHZurich}}$\orcidlink{0000-0001-8823-479X}}, 
  \mbox{J.~D\"orner$^{\ref{AFFIL::GermanyUBochum}}$\orcidlink{0000-0001-6692-6293}}, 
  \mbox{M.~Doro$^{\ref{AFFIL::ItalyUPadovaandINFN}}$\orcidlink{0000-0001-9104-3214}}, 
  \mbox{R.~D.~C.~dos~Anjos$^{\ref{AFFIL::BrazilUFPR}}$\orcidlink{0000-0002-6463-2272}}, 
  \mbox{J.-L.~Dournaux$^{\ref{AFFIL::FranceObservatoiredeParis}}$}, 
  \mbox{D.~Dravins$^{\ref{AFFIL::SwedenLundObservatory}}$\orcidlink{0000-0001-9024-0400}}, 
  \mbox{C.~Duangchan$^{\ref{AFFIL::GermanyUErlangenECAP},\ref{AFFIL::ThailandNARIT}}$\orcidlink{0009-0003-8227-6552}}, 
  \mbox{C.~Dubos$^{\ref{AFFIL::FranceIJCLab}}$}, 
  \mbox{L.~Ducci$^{\ref{AFFIL::GermanyIAAT}}$}, 
  \mbox{V.~V.~Dwarkadas$^{\ref{AFFIL::USAUChicagoDAA}}$\orcidlink{0000-0002-4661-7001}}, 
  \mbox{J.~Ebr$^{\ref{AFFIL::CzechRepublicFZU}}$}, 
  \mbox{C.~Eckner$^{\ref{AFFIL::FranceLAPPUSavoieMontBlanc},\ref{AFFIL::FranceLAPTh}}$}, 
  \mbox{K.~Egberts$^{\ref{AFFIL::GermanyUPotsdam}}$}, 
  \mbox{S.~Einecke$^{\ref{AFFIL::AustraliaUAdelaide}}$\orcidlink{0000-0001-9687-8237}}, 
  \mbox{D.~Els\"asser$^{\ref{AFFIL::GermanyUDortmundTU}}$\orcidlink{0000-0001-6796-3205}}, 
  \mbox{G.~Emery$^{\ref{AFFIL::FranceCPPMUAixMarseille}}$\orcidlink{0000-0001-6155-4742}}, 
  \mbox{M.~Escobar~Godoy$^{\ref{AFFIL::USASCIPP}}$}, 
  \mbox{J.~Escudero$^{\ref{AFFIL::SpainIAACSIC}}$\orcidlink{0000-0002-4131-655X}}, 
  \mbox{P.~Esposito$^{\ref{AFFIL::ItalyIUSSPaviaINAF},\ref{AFFIL::ItalyIASFMilano}}$\orcidlink{0000-0003-4849-5092}}, 
  \mbox{D.~Falceta-Goncalves$^{\ref{AFFIL::BrazilEACHUSaoPaulo}}$\orcidlink{0000-0002-1914-6654}}, 
  \mbox{V.~Fallah~Ramazani$^{\ref{AFFIL::GermanyUBochum}}$\orcidlink{0000-0001-8991-7744}}, 
  \mbox{A.~Faure$^{\ref{AFFIL::FranceLUPMUMontpellier}}$}, 
  \mbox{E.~Fedorova$^{\ref{AFFIL::ItalyORoma},\ref{AFFIL::UkraineAstObsofUKyiv}}$\orcidlink{0000-0002-8882-7496}}, 
  \mbox{S.~Fegan$^{\ref{AFFIL::FranceLLREcolePolytechnique}}$\orcidlink{0000-0002-9978-2510}}, 
  \mbox{K.~Feijen$^{\ref{AFFIL::FranceAPCUParisCite}}$\orcidlink{0000-0003-1476-3714}}, 
  \mbox{Q.~Feng$^{\ref{AFFIL::USACfAHarvardSmithsonian}}$}, 
  \mbox{G.~Ferrand$^{\ref{AFFIL::CanadaUManitoba},\ref{AFFIL::JapanRIKEN}}$\orcidlink{0000-0002-4231-8717}}, 
  \mbox{F.~Ferrarotto$^{\ref{AFFIL::ItalyINFNRomaLaSapienza}}$\orcidlink{0000-0001-5464-0378}}, 
  \mbox{E.~Fiandrini$^{\ref{AFFIL::ItalyUPerugiaandINFN}}$}, 
  \mbox{A.~Fiasson$^{\ref{AFFIL::FranceLAPPUSavoieMontBlanc}}$}, 
  \mbox{V.~Fioretti$^{\ref{AFFIL::ItalyOASBologna}}$\orcidlink{0000-0002-6082-5384}}, 
  \mbox{L.~Foffano$^{\ref{AFFIL::ItalyIAPS}}$\orcidlink{0000-0002-0709-9707}}, 
  \mbox{L.~Font~Guiteras$^{\ref{AFFIL::SpainUABandCERESIEEC}}$\orcidlink{0000-0003-2109-5961}}, 
  \mbox{G.~Fontaine$^{\ref{AFFIL::FranceLLREcolePolytechnique}}$\orcidlink{0000-0002-6443-5025}}, 
  \mbox{S.~Fr\"ose$^{\ref{AFFIL::GermanyUDortmundTU}}$\orcidlink{0000-0003-1832-4129}}, 
  \mbox{S.~Fukami$^{\ref{AFFIL::SwitzerlandETHZurich}}$}, 
  \mbox{Y.~Fukui$^{\ref{AFFIL::JapanUNagoya}}$\orcidlink{0000-0002-8966-9856}}, 
  \mbox{S.~Funk$^{\ref{AFFIL::GermanyUErlangenECAP}}$\orcidlink{0000-0002-2012-0080}}, 
  \mbox{D.~Gaggero$^{\ref{AFFIL::ItalyINFNPisa}}$}, 
  \mbox{G.~Galanti$^{\ref{AFFIL::ItalyIASFMilano}}$\orcidlink{0000-0001-7254-3029}}, 
  \mbox{G.~Galaz$^{\ref{AFFIL::ChileUPontificiaCatolicadeChile}}$\orcidlink{0000-0002-8835-0739}}, 
  \mbox{Y.~A.~Gallant$^{\ref{AFFIL::FranceLUPMUMontpellier}}$}, 
  \mbox{S.~Gallozzi$^{\ref{AFFIL::ItalyORoma}}$\orcidlink{0000-0003-4456-9875}}, 
  \mbox{V.~Gammaldi$^{\ref{AFFIL::SpainIFTUAMCSIC}}$\orcidlink{0000-0003-1826-6117}}, 
  \mbox{C.~Gasbarra$^{\ref{AFFIL::ItalyINFNRomaTorVergata}}$\orcidlink{0000-0001-8335-9614}}, 
  \mbox{M.~Gaug$^{\ref{AFFIL::SpainUABandCERESIEEC}}$\orcidlink{0000-0001-8442-7877}}, 
  \mbox{A.~Ghalumyan$^{\ref{AFFIL::ArmeniaNSLAlikhanyan}}$}, 
  \mbox{F.~Gianotti$^{\ref{AFFIL::ItalyOASBologna}}$\orcidlink{0000-0003-4666-119X}}, 
  \mbox{M.~Giarrusso$^{\ref{AFFIL::ItalyINFNCatania}}$}, 
  \mbox{N.~Giglietto$^{\ref{AFFIL::ItalyPolitecnicoBari},\ref{AFFIL::ItalyINFNBari}}$\orcidlink{0000-0002-9021-2888}}, 
  \mbox{F.~Giordano$^{\ref{AFFIL::ItalyUandINFNBari}}$\orcidlink{0000-0002-8651-2394}}, 
  \mbox{A.~Giuliani$^{\ref{AFFIL::ItalyIASFMilano}}$}, 
  \mbox{J.-F.~Glicenstein$^{\ref{AFFIL::FranceCEAIRFUDPhP}}$}, 
  \mbox{J.~Glombitza$^{\ref{AFFIL::GermanyUErlangenECAP}}$}, 
  \mbox{P.~Goldoni$^{\ref{AFFIL::FranceAPCUParisCiteCEAaffiliatedpersonnel}}$\orcidlink{0000-0001-5638-5817}}, 
  \mbox{J.~M.~Gonz\'alez$^{\ref{AFFIL::ChileUAndresBello}}$\orcidlink{0000-0002-2413-0681}}, 
  \mbox{M.~M.~Gonz\'alez$^{\ref{AFFIL::MexicoUNAMMexico}}$}, 
  \mbox{J.~Goulart~Coelho$^{\ref{AFFIL::BrazilUFES}}$\orcidlink{0000-0001-9386-1042}}, 
  \mbox{J.~Granot$^{\ref{AFFIL::IsraelOpenUniversityofIsrael},\ref{AFFIL::USAGWUWashingtonDC}}$}, 
  \mbox{D.~Grasso$^{\ref{AFFIL::ItalyINFNPisa}}$}, 
  \mbox{R.~Grau$^{\ref{AFFIL::SpainIFAEBIST}}$\orcidlink{0000-0002-1891-6290}}, 
  \mbox{D.~Green$^{\ref{AFFIL::GermanyMPP}}$\orcidlink{0000-0003-0768-2203}}, 
  \mbox{J.~G.~Green$^{\ref{AFFIL::GermanyMPP}}$\orcidlink{0000-0002-1130-6692}}, 
  \mbox{T.~Greenshaw$^{\ref{AFFIL::UnitedKingdomULiverpool}}$}, 
  \mbox{G.~Grolleron$^{\ref{AFFIL::FranceLPNHEUSorbonne}}$}, 
  \mbox{J.~Grube$^{\ref{AFFIL::UnitedKingdomKingsCollege}}$}, 
  \mbox{O.~Gueta$^{\ref{AFFIL::GermanyDESY}}$\orcidlink{0000-0002-9440-2398}}, 
  \mbox{S.~Gunji$^{\ref{AFFIL::JapanUYamagata}}$\orcidlink{0000-0002-5881-2445}}, 
  \mbox{D.~Hadasch$^{\ref{AFFIL::JapanUTokyoICRR}}$\orcidlink{0000-0001-8663-6461}}, 
  \mbox{P.~Hamal$^{\ref{AFFIL::CzechRepublicFZU}}$\orcidlink{0000-0003-3139-7234}}, 
  \mbox{W.~Hanlon$^{\ref{AFFIL::USACfAHarvardSmithsonian}}$\orcidlink{0000-0002-0109-4737}}, 
  \mbox{S.~Hara$^{\ref{AFFIL::JapanUYamanashiGakuin}}$\orcidlink{0009-0001-1220-7717}}, 
  \mbox{V.~M.~Harvey$^{\ref{AFFIL::AustraliaUAdelaide}}$\orcidlink{0000-0001-9090-8415}}, 
  \mbox{K.~Hashiyama$^{\ref{AFFIL::JapanUTokyoICRR}}$}, 
  \mbox{T.~Hassan$^{\ref{AFFIL::SpainCIEMAT}}$\orcidlink{0000-0002-4758-9196}}, 
  \mbox{M.~Heller$^{\ref{AFFIL::SwitzerlandUGenevaDPNC}}$}, 
  \mbox{S.~Hern\'andez~Cadena$^{\ref{AFFIL::MexicoUNAMMexico}}$\orcidlink{0000-0002-2565-8365}}, 
  \mbox{J.~Hie$^{\ref{AFFIL::FranceIRAPUToulouse}}$}, 
  \mbox{N.~Hiroshima$^{\ref{AFFIL::JapanUTokyoICRR}}$}, 
  \mbox{B.~Hnatyk$^{\ref{AFFIL::UkraineAstObsofUKyiv}}$\orcidlink{0000-0001-7113-4709}}, 
  \mbox{R.~Hnatyk$^{\ref{AFFIL::UkraineAstObsofUKyiv}}$}, 
  \mbox{D.~Hoffmann$^{\ref{AFFIL::FranceCPPMUAixMarseille}}$\orcidlink{0000-0001-5209-5265}}, 
  \mbox{W.~Hofmann$^{\ref{AFFIL::GermanyMPIK}}$}, 
  \mbox{M.~Holler$^{\ref{AFFIL::AustriaUInnsbruck}}$}, 
  \mbox{D.~Horan$^{\ref{AFFIL::FranceLLREcolePolytechnique}}$}, 
  \mbox{P.~Horvath$^{\ref{AFFIL::CzechRepublicUOlomouc}}$\orcidlink{0000-0002-6710-5339}}, 
  \mbox{T.~Hovatta$^{\ref{AFFIL::FinlandUTurku}}$}, 
  \mbox{D.~Hrupec$^{\ref{AFFIL::CroatiaUOsijek}}$\orcidlink{0000-0002-7027-5021}}, 
  \mbox{S.~Hussain$^{\ref{AFFIL::BrazilIAGUSaoPaulo},\ref{AFFIL::ItalyGSSIandINFNAquila}}$\orcidlink{0000-0002-0458-0490}}, 
  \mbox{M.~Iarlori$^{\ref{AFFIL::ItalyUandINFNAquila}}$}, 
  \mbox{T.~Inada$^{\ref{AFFIL::JapanUTokyoICRR}}$\orcidlink{0000-0002-6923-9314}}, 
  \mbox{F.~Incardona$^{\ref{AFFIL::ItalyOCatania}}$}, 
  \mbox{Y.~Inome$^{\ref{AFFIL::JapanUTokyoICRR}}$}, 
  \mbox{S.~Inoue$^{\ref{AFFIL::JapanRIKEN}}$}, 
  \mbox{F.~Iocco$^{\ref{AFFIL::ItalyUNapoli},\ref{AFFIL::ItalyINFNNapoli}}$}, 
  \mbox{K.~Ishio$^{\ref{AFFIL::PolandULodz}}$}, 
  \mbox{M.~Jamrozy$^{\ref{AFFIL::PolandUJagiellonian}}$\orcidlink{0000-0002-0870-7778}}, 
  \mbox{P.~Janecek$^{\ref{AFFIL::CzechRepublicFZU}}$}, 
  \mbox{F.~Jankowsky$^{\ref{AFFIL::GermanyLSW}}$}, 
  \mbox{C.~Jarnot$^{\ref{AFFIL::FranceIRAPUToulouse}}$}, 
  \mbox{P.~Jean$^{\ref{AFFIL::FranceIRAPUToulouse}}$\orcidlink{0000-0002-1757-9560}}, 
  \mbox{I.~Jim\'enez~Mart{\'\i}nez$^{\ref{AFFIL::SpainCIEMAT}}$\orcidlink{0000-0003-2150-6919}}, 
  \mbox{W.~Jin$^{\ref{AFFIL::USAUAlabamaTuscaloosa}}$\orcidlink{0000-0002-1089-1754}}, 
  \mbox{L.~Jocou$^{\ref{AFFIL::FranceIPAGUGrenobleAlpes}}$}, 
  \mbox{C.~Juramy-Gilles$^{\ref{AFFIL::FranceLPNHEUSorbonne}}$\orcidlink{0000-0002-3145-9258}}, 
  \mbox{J.~Jurysek$^{\ref{AFFIL::CzechRepublicFZU}}$\orcidlink{0000-0002-3130-4168}}, 
  \mbox{O.~Kalekin$^{\ref{AFFIL::GermanyUErlangenECAP}}$}, 
  \mbox{D.~Kantzas$^{\ref{AFFIL::FranceLAPTh}}$\orcidlink{0000-0002-7364-606X}}, 
  \mbox{V.~Karas$^{\ref{AFFIL::CzechRepublicASU}}$}, 
  \mbox{S.~Kaufmann$^{\ref{AFFIL::UnitedKingdomUDurham}}$}, 
  \mbox{D.~Kerszberg$^{\ref{AFFIL::SpainIFAEBIST}}$\orcidlink{0000-0002-5289-1509}}, 
  \mbox{B.~Kh\'elifi$^{\ref{AFFIL::FranceAPCUParisCite}}$\orcidlink{0000-0001-6876-5577}}, 
  \mbox{D.~B.~Kieda$^{\ref{AFFIL::USAUUtah}}$\orcidlink{0000-0003-4785-0101}}, 
  \mbox{T.~Kleiner$^{\ref{AFFIL::GermanyDESY}}$\orcidlink{0000-0002-4260-9186}}, 
  \mbox{W.~Klu\'zniak$^{\ref{AFFIL::PolandNicolausCopernicusAstronomicalCenter}}$}, 
  \mbox{Y.~Kobayashi$^{\ref{AFFIL::JapanUTokyoICRR}}$}, 
  \mbox{K.~Kohri$^{\ref{AFFIL::JapanKEK}}$}, 
  \mbox{N.~Komin$^{\ref{AFFIL::SouthAfricaUWitwatersrand}}$\orcidlink{0000-0003-3280-0582}}, 
  \mbox{P.~Kornecki$^{\ref{AFFIL::FranceObservatoiredeParis}}$\orcidlink{0000-0002-2706-7438}}, 
  \mbox{K.~Kosack$^{\ref{AFFIL::FranceCEAIRFUDAp}}$\orcidlink{0000-0001-8424-3621}}, 
  \mbox{H.~Kubo$^{\ref{AFFIL::JapanUTokyoICRR}}$\orcidlink{0000-0001-9159-9853}}, 
  \mbox{J.~Kushida$^{\ref{AFFIL::JapanUTokai}}$\orcidlink{0000-0002-8002-8585}}, 
  \mbox{A.~La~Barbera$^{\ref{AFFIL::ItalyIASFPalermo}}$\orcidlink{0000-0002-5880-8913}}, 
  \mbox{N.~La~Palombara$^{\ref{AFFIL::ItalyIASFMilano}}$\orcidlink{0000-0001-7015-6359}}, 
  \mbox{M.~L\'ainez$^{\ref{AFFIL::SpainUCMAltasEnergias}}$\orcidlink{0000-0003-3848-922X}}, 
  \mbox{A.~Lamastra$^{\ref{AFFIL::ItalyORoma}}$\orcidlink{0000-0003-2403-913X}}, 
  \mbox{J.~Lapington$^{\ref{AFFIL::UnitedKingdomULeicester}}$}, 
  \mbox{S.~Lazarevi\'c$^{\ref{AFFIL::AustraliaUWesternSydney}}$\orcidlink{0000-0001-6109-8548}}, 
  \mbox{J.~Lazendic-Galloway$^{\ref{AFFIL::AustraliaUMonash}}$}, 
  \mbox{S.~Leach$^{\ref{AFFIL::UnitedKingdomULeicester}}$\orcidlink{0000-0003-2129-3175}}, 
  \mbox{M.~Lemoine-Goumard$^{\ref{AFFIL::FranceLP2IUBordeaux}}$}, 
  \mbox{J.-P.~Lenain$^{\ref{AFFIL::FranceLPNHEUSorbonne}}$\orcidlink{0000-0001-7284-9220}}, 
  \mbox{G.~Leto$^{\ref{AFFIL::ItalyOCatania}}$\orcidlink{0000-0002-0040-5011}}, 
  \mbox{F.~Leuschner$^{\ref{AFFIL::GermanyIAAT}}$\orcidlink{0000-0001-9037-0272}}, 
  \mbox{E.~Lindfors$^{\ref{AFFIL::FinlandUTurku}}$}, 
  \mbox{M.~Linhoff$^{\ref{AFFIL::GermanyUDortmundTU}}$\orcidlink{0000-0001-7993-8189}}, 
  \mbox{I.~Liodakis$^{\ref{AFFIL::FinlandUTurku}}$\orcidlink{0000-0001-9200-4006}}, 
  \mbox{L.~Lo{\"\i}c$^{\ref{AFFIL::FranceCEAIRFUDPhP}}$}, 
  \mbox{S.~Lombardi$^{\ref{AFFIL::ItalyORoma}}$\orcidlink{0000-0002-6336-865X}}, 
  \mbox{F.~Longo$^{\ref{AFFIL::ItalyUandINFNTrieste}}$\orcidlink{0000-0003-2501-2270}}, 
  \mbox{R.~L\'opez-Coto$^{\ref{AFFIL::SpainIAACSIC}}$}, 
  \mbox{M.~L\'opez-Moya$^{\ref{AFFIL::SpainUCMAltasEnergias}}$\orcidlink{0000-0002-8791-7908}}, 
  \mbox{A.~L\'opez-Oramas$^{\ref{AFFIL::SpainIAC}}$\orcidlink{0000-0003-4603-1884}}, 
  \mbox{S.~Loporchio$^{\ref{AFFIL::ItalyPolitecnicoBari},\ref{AFFIL::ItalyINFNBari}}$}, 
  \mbox{J.~Lozano~Bahilo$^{\ref{AFFIL::SpainUAlcala}}$\orcidlink{0000-0003-0613-140X}}, 
  \mbox{P.~L.~Luque-Escamilla$^{\ref{AFFIL::SpainUJaen}}$}, 
  \mbox{O.~Macias$^{\ref{AFFIL::NetherlandsUAmsterdam}}$\orcidlink{0000-0001-8867-2693}}, 
  \mbox{G.~Maier$^{\ref{AFFIL::GermanyDESY}}$\orcidlink{0000-0001-9868-4700}}, 
  \mbox{P.~Majumdar$^{\ref{AFFIL::IndiaSahaInstitute}}$\orcidlink{0000-0002-5481-5040}}, 
  \mbox{D.~Malyshev$^{\ref{AFFIL::GermanyIAAT}}$\orcidlink{0000-0001-9689-2194}}, 
  \mbox{D.~Malyshev$^{\ref{AFFIL::GermanyUErlangenECAP}}$\orcidlink{0000-0002-9102-4854}}, 
  \mbox{D.~Mandat$^{\ref{AFFIL::CzechRepublicFZU}}$}, 
  \mbox{G.~Manic\`o$^{\ref{AFFIL::ItalyINFNCatania},\ref{AFFIL::ItalyUCatania}}$}, 
  \mbox{P.~Marinos$^{\ref{AFFIL::AustraliaUAdelaide}}$\orcidlink{0000-0003-1734-0215}}, 
  \mbox{S.~Markoff$^{\ref{AFFIL::NetherlandsUAmsterdam}}$\orcidlink{0000-0001-9564-0876}}, 
  \mbox{I.~M\'arquez$^{\ref{AFFIL::SpainIAACSIC}}$\orcidlink{0000-0003-2629-1945}}, 
  \mbox{P.~Marquez$^{\ref{AFFIL::SpainIFAEBIST}}$\orcidlink{0000-0002-9591-7967}}, 
  \mbox{G.~Marsella$^{\ref{AFFIL::ItalyUPalermo},\ref{AFFIL::ItalyINFNCatania}}$\orcidlink{0000-0002-3152-8874}}, 
  \mbox{J.~Mart{\'\i}$^{\ref{AFFIL::SpainUJaen}}$}, 
  \mbox{P.~Martin$^{\ref{AFFIL::FranceIRAPUToulouse}}$\orcidlink{0000-0002-7670-6320}}, 
  \mbox{G.~A.~Mart{\'\i}nez$^{\ref{AFFIL::SpainCIEMAT}}$\orcidlink{0000-0002-1061-8520}}, 
  \mbox{M.~Mart{\'\i}nez$^{\ref{AFFIL::SpainIFAEBIST}}$}, 
  \mbox{O.~Martinez$^{\ref{AFFIL::SpainUCMElectronica},\ref{AFFIL::SpainUPCMadrid}}$\orcidlink{0000-0002-3353-7707}}, 
  \mbox{C.~Marty$^{\ref{AFFIL::FranceIRAPUToulouse}}$}, 
  \mbox{A.~Mas-Aguilar$^{\ref{AFFIL::SpainUCMAltasEnergias}}$\orcidlink{0000-0002-8893-9009}}, 
  \mbox{M.~Mastropietro$^{\ref{AFFIL::ItalyORoma}}$\orcidlink{0000-0002-6324-5713}}, 
  \mbox{G.~Maurin$^{\ref{AFFIL::FranceLAPPUSavoieMontBlanc}}$}, 
  \mbox{W.~Max-Moerbeck$^{\ref{AFFIL::ChileUdeChile}}$\orcidlink{0000-0002-5491-5244}}, 
  \mbox{D.~Mazin$^{\ref{AFFIL::JapanUTokyoICRR},\ref{AFFIL::GermanyMPP}}$}, 
  \mbox{D.~Melkumyan$^{\ref{AFFIL::GermanyDESY}}$}, 
  \mbox{S.~Menchiari$^{\ref{AFFIL::ItalyOArcetri},\ref{AFFIL::ItalyINFNPisa}}$}, 
  \mbox{E.~Mestre$^{\ref{AFFIL::SpainICECSIC}}$}, 
  \mbox{J.-L.~Meunier$^{\ref{AFFIL::FranceLPNHEUSorbonne}}$}, 
  \mbox{D.~M.-A.~Meyer$^{\ref{AFFIL::GermanyUPotsdam}}$\orcidlink{0000-0001-8258-9813}}, 
  \mbox{D.~Miceli$^{\ref{AFFIL::ItalyINFNPadova}}$\orcidlink{0000-0002-2686-0098}}, 
  \mbox{M.~Michailidis$^{\ref{AFFIL::GermanyIAAT}}$}, 
  \mbox{J.~Micha{\l}owski$^{\ref{AFFIL::PolandIFJ}}$}, 
  \mbox{T.~Miener$^{\ref{AFFIL::SpainUCMAltasEnergias}}$}, 
  \mbox{J.~M.~Miranda$^{\ref{AFFIL::SpainUCMElectronica},\ref{AFFIL::SpainIPARCOSInstitute}}$}, 
  \mbox{A.~Mitchell$^{\ref{AFFIL::GermanyUErlangenECAP}}$\orcidlink{0000-0003-3631-5648}}, 
  \mbox{M.~Mizote$^{\ref{AFFIL::JapanUKonan}}$}, 
  \mbox{T.~Mizuno$^{\ref{AFFIL::JapanHASC}}$}, 
  \mbox{R.~Moderski$^{\ref{AFFIL::PolandNicolausCopernicusAstronomicalCenter}}$\orcidlink{0000-0002-8663-3882}}, 
  \mbox{L.~Mohrmann$^{\ref{AFFIL::GermanyMPIK}}$\orcidlink{0000-0002-9667-8654}}, 
  \mbox{M.~Molero$^{\ref{AFFIL::SpainIAC}}$\orcidlink{0000-0003-0967-715X}}, 
  \mbox{C.~Molfese$^{\ref{AFFIL::ItalyINAF}}$\orcidlink{0000-0002-2756-9075}}, 
  \mbox{E.~Molina$^{\ref{AFFIL::SpainIAC}}$\orcidlink{0000-0003-1204-5516}}, 
  \mbox{T.~Montaruli$^{\ref{AFFIL::SwitzerlandUGenevaDPNC}}$}, 
  \mbox{A.~Moralejo$^{\ref{AFFIL::SpainIFAEBIST}}$}, 
  \mbox{D.~Morcuende$^{\ref{AFFIL::SpainUCMAltasEnergias},\ref{AFFIL::SpainIAACSIC}}$\orcidlink{0000-0001-9400-0922}}, 
  \mbox{K.~Morik$^{\ref{AFFIL::GermanyUDortmundTU}}$\orcidlink{0000-0003-1153-5986}}, 
  \mbox{A.~Morselli$^{\ref{AFFIL::ItalyINFNRomaTorVergata}}$\orcidlink{0000-0002-7704-9553}}, 
  \mbox{E.~Moulin$^{\ref{AFFIL::FranceCEAIRFUDPhP}}$\orcidlink{0000-0003-4007-0145}}, 
  \mbox{V.~Moya~Zamanillo$^{\ref{AFFIL::SpainUCMAltasEnergias}}$\orcidlink{0000-0001-9407-5545}}, 
  \mbox{R.~Mukherjee$^{\ref{AFFIL::USABarnardCollegeColumbiaUniversity}}$\orcidlink{0000-0002-3223-0754}}, 
  \mbox{K.~Munari$^{\ref{AFFIL::ItalyOCatania}}$}, 
  \mbox{A.~Muraczewski$^{\ref{AFFIL::PolandNicolausCopernicusAstronomicalCenter}}$}, 
  \mbox{H.~Muraishi$^{\ref{AFFIL::JapanUKitasato}}$\orcidlink{0000-0003-3054-5725}}, 
  \mbox{T.~Nakamori$^{\ref{AFFIL::JapanUYamagata}}$\orcidlink{0000-0002-7308-2356}}, 
  \mbox{L.~Nava$^{\ref{AFFIL::ItalyOBrera}}$\orcidlink{0000-0001-5960-0455}}, 
  \mbox{A.~Nayak$^{\ref{AFFIL::UnitedKingdomUDurham}}$}, 
  \mbox{R.~Nemmen$^{\ref{AFFIL::BrazilIAGUSaoPaulo},\ref{AFFIL::USAStanford}}$\orcidlink{0000-0003-3956-0331}}, 
  \mbox{L.~Nickel$^{\ref{AFFIL::GermanyUDortmundTU}}$\orcidlink{0000-0001-7110-0533}}, 
  \mbox{J.~Niemiec$^{\ref{AFFIL::PolandIFJ}}$\orcidlink{0000-0001-6036-8569}}, 
  \mbox{D.~Nieto$^{\ref{AFFIL::SpainUCMAltasEnergias}}$\orcidlink{0000-0003-3343-0755}}, 
  \mbox{M.~Nievas~Rosillo$^{\ref{AFFIL::SpainIAC}}$\orcidlink{0000-0002-8321-9168}}, 
  \mbox{M.~Niko{\l}ajuk$^{\ref{AFFIL::PolandUBiaystok}}$\orcidlink{0000-0003-4075-6745}}, 
  \mbox{K.~Nishijima$^{\ref{AFFIL::JapanUTokai}}$\orcidlink{0000-0002-1830-4251}}, 
  \mbox{K.~Noda$^{\ref{AFFIL::JapanUTokyoICRR}}$\orcidlink{0000-0003-1397-6478}}, 
  \mbox{D.~Nosek$^{\ref{AFFIL::CzechRepublicUPrague}}$\orcidlink{0000-0001-6219-200X}}, 
  \mbox{B.~Novosyadlyj$^{\ref{AFFIL::UkraineAstObsofULviv}}$}, 
  \mbox{V.~Novotny$^{\ref{AFFIL::CzechRepublicUPrague}}$\orcidlink{0000-0002-4319-4541}}, 
  \mbox{S.~Nozaki$^{\ref{AFFIL::GermanyMPP}}$\orcidlink{0000-0002-6246-2767}}, 
  \mbox{P.~O'Brien$^{\ref{AFFIL::UnitedKingdomULeicester}}$\orcidlink{0000-0002-5128-1899}}, 
  \mbox{M.~Ohishi$^{\ref{AFFIL::JapanUTokyoICRR}}$\orcidlink{0000-0002-5056-0968}}, 
  \mbox{Y.~Ohtani$^{\ref{AFFIL::JapanUTokyoICRR}}$\orcidlink{0000-0001-7042-4958}}, 
  \mbox{A.~Okumura$^{\ref{AFFIL::JapanUNagoyaISEE},\ref{AFFIL::JapanUNagoyaKMI}}$\orcidlink{0000-0002-3055-7964}}, 
  \mbox{J.-F.~Olive$^{\ref{AFFIL::FranceIRAPUToulouse}}$}, 
  \mbox{B.~Olmi$^{\ref{AFFIL::ItalyOPalermo},\ref{AFFIL::ItalyOArcetri}}$}, 
  \mbox{R.~A.~Ong$^{\ref{AFFIL::USAUCLA}}$\orcidlink{0000-0002-4837-5253}}, 
  \mbox{M.~Orienti$^{\ref{AFFIL::ItalyRadioastronomiaINAF}}$\orcidlink{0000-0003-4470-7094}}, 
  \mbox{R.~Orito$^{\ref{AFFIL::JapanUTokushima}}$}, 
  \mbox{M.~Orlandini$^{\ref{AFFIL::ItalyOASBologna}}$\orcidlink{0000-0003-0946-3151}}, 
  \mbox{E.~Orlando$^{\ref{AFFIL::ItalyUandINFNTrieste}}$}, 
  \mbox{M.~Ostrowski$^{\ref{AFFIL::PolandUJagiellonian}}$\orcidlink{0000-0002-9199-7031}}, 
  \mbox{N.~Otte$^{\ref{AFFIL::USAGeorgiaTech}}$\orcidlink{0000-0002-5955-6383}}, 
  \mbox{I.~Oya$^{\ref{AFFIL::GermanyCTAOHeidelberg}}$\orcidlink{0000-0002-3881-9324}}, 
  \mbox{I.~Pagano$^{\ref{AFFIL::ItalyOCatania}}$\orcidlink{0000-0001-9573-4928}}, 
  \mbox{A.~Pagliaro$^{\ref{AFFIL::ItalyIASFPalermo}}$\orcidlink{0000-0002-6841-1362}}, 
  \mbox{M.~Palatiello$^{\ref{AFFIL::ItalyUUdineandINFNTrieste}}$}, 
  \mbox{G.~Panebianco$^{\ref{AFFIL::ItalyOASBologna}}$\orcidlink{0000-0002-3410-8613}}, 
  \mbox{J.~M.~Paredes$^{\ref{AFFIL::SpainICCUB}}$}, 
  \mbox{N.~Parmiggiani$^{\ref{AFFIL::ItalyOASBologna}}$\orcidlink{0000-0002-4535-5329}}, 
  \mbox{S.~R.~Patel$^{\ref{AFFIL::FranceIJCLab}}$\orcidlink{0000-0001-8965-7292}}, 
  \mbox{B.~Patricelli$^{\ref{AFFIL::ItalyORoma},\ref{AFFIL::ItalyUPisa}}$\orcidlink{0000-0001-6709-0969}}, 
  \mbox{D.~Pavlovi\'c$^{\ref{AFFIL::CroatiaURijeka}}$}, 
  \mbox{A.~Pe'er$^{\ref{AFFIL::GermanyMPP}}$\orcidlink{0000-0001-8667-0889}}, 
  \mbox{M.~Pech$^{\ref{AFFIL::CzechRepublicFZU}}$}, 
  \mbox{M.~Pecimotika$^{\ref{AFFIL::CroatiaURijeka},\ref{AFFIL::CroatiaIRB}}$\orcidlink{0000-0002-4699-1845}}, 
  \mbox{M.~Peresano$^{\ref{AFFIL::ItalyUTorino},\ref{AFFIL::ItalyINFNTorino}}$\orcidlink{0000-0002-7537-7334}}, 
  \mbox{J.~P\'erez-Romero$^{\ref{AFFIL::SpainIFTUAMCSIC},\ref{AFFIL::SloveniaUNovaGoricaCAC}}$\orcidlink{0000-0002-9408-3120}}, 
  \mbox{G.~Peron$^{\ref{AFFIL::FranceAPCUParisCite}}$}, 
  \mbox{M.~Persic$^{\ref{AFFIL::ItalyOPadova},\ref{AFFIL::ItalyOandINFNTrieste}}$\orcidlink{0000-0003-1853-4900}}, 
  \mbox{P.-O.~Petrucci$^{\ref{AFFIL::FranceIPAGUGrenobleAlpes}}$\orcidlink{0000-0001-6061-3480}}, 
  \mbox{O.~Petruk$^{\ref{AFFIL::UkraineIAPMMLviv}}$\orcidlink{0000-0003-3487-0349}}, 
  \mbox{F.~Pfeifle$^{\ref{AFFIL::GermanyUWurzburg}}$}, 
  \mbox{F.~Pintore$^{\ref{AFFIL::ItalyIASFPalermo}}$\orcidlink{0000-0002-3869-2925}}, 
  \mbox{G.~Pirola$^{\ref{AFFIL::GermanyMPP}}$}, 
  \mbox{C.~Pittori$^{\ref{AFFIL::ItalyORoma}}$\orcidlink{0000-0001-6661-9779}}, 
  \mbox{C.~Plard$^{\ref{AFFIL::FranceLAPPUSavoieMontBlanc}}$\orcidlink{0000-0002-4061-3800}}, 
  \mbox{F.~Podobnik$^{\ref{AFFIL::ItalyUSienaandINFN}}$\orcidlink{0000-0001-6125-9487}}, 
  \mbox{M.~Pohl$^{\ref{AFFIL::GermanyUPotsdam},\ref{AFFIL::GermanyDESY}}$\orcidlink{0000-0001-7861-1707}}, 
  \mbox{E.~Pons$^{\ref{AFFIL::FranceLAPPUSavoieMontBlanc}}$\orcidlink{0000-0002-7601-9811}}, 
  \mbox{E.~Prandini$^{\ref{AFFIL::ItalyUPadovaandINFN}}$\orcidlink{0000-0003-4502-9053}}, 
  \mbox{J.~Prast$^{\ref{AFFIL::FranceLAPPUSavoieMontBlanc}}$}, 
  \mbox{G.~Principe$^{\ref{AFFIL::ItalyUandINFNTrieste}}$}, 
  \mbox{C.~Priyadarshi$^{\ref{AFFIL::SpainIFAEBIST}}$\orcidlink{0000-0002-9160-9617}}, 
  \mbox{N.~Produit$^{\ref{AFFIL::SwitzerlandUGenevaISDC}}$\orcidlink{0000-0001-7138-7677}}, 
  \mbox{D.~Prokhorov$^{\ref{AFFIL::NetherlandsUAmsterdam}}$}, 
  \mbox{E.~Pueschel$^{\ref{AFFIL::GermanyDESY}}$\orcidlink{0000-0002-0529-1973}}, 
  \mbox{G.~P\"uhlhofer$^{\ref{AFFIL::GermanyIAAT}}$}, 
  \mbox{M.~L.~Pumo$^{\ref{AFFIL::ItalyUCatania},\ref{AFFIL::ItalyINFNCatania}}$}, 
  \mbox{M.~Punch$^{\ref{AFFIL::FranceAPCUParisCite}}$\orcidlink{0000-0002-4710-2165}}, 
  \mbox{A.~Quirrenbach$^{\ref{AFFIL::GermanyLSW}}$}, 
  \mbox{S.~Rain\`o$^{\ref{AFFIL::ItalyUandINFNBari}}$\orcidlink{0000-0002-9181-0345}}, 
  \mbox{N.~Randazzo$^{\ref{AFFIL::ItalyINFNCatania}}$}, 
  \mbox{R.~Rando$^{\ref{AFFIL::ItalyUPadovaandINFN}}$\orcidlink{0000-0001-6992-818X}}, 
  \mbox{T.~Ravel$^{\ref{AFFIL::FranceIRAPUToulouse}}$}, 
  \mbox{S.~Razzaque$^{\ref{AFFIL::SouthAfricaUJohannesburg},\ref{AFFIL::USAGWUWashingtonDC}}$\orcidlink{0000-0002-0130-2460}}, 
  \mbox{M.~Regeard$^{\ref{AFFIL::FranceAPCUParisCite}}$\orcidlink{0000-0002-3844-6003}}, 
  \mbox{P.~Reichherzer$^{\ref{AFFIL::UnitedKingdomUOxford},\ref{AFFIL::GermanyUBochum}}$\orcidlink{0000-0003-4513-8241}}, 
  \mbox{A.~Reimer$^{\ref{AFFIL::AustriaUInnsbruck}}$\orcidlink{0000-0001-8604-7077}}, 
  \mbox{O.~Reimer$^{\ref{AFFIL::AustriaUInnsbruck}}$\orcidlink{0000-0001-6953-1385}}, 
  \mbox{A.~Reisenegger$^{\ref{AFFIL::ChileUPontificiaCatolicadeChile},\ref{AFFIL::ChileUMCE}}$\orcidlink{0000-0003-4059-6796}}, 
  \mbox{T.~Reposeur$^{\ref{AFFIL::FranceLP2IUBordeaux}}$}, 
  \mbox{B.~Reville$^{\ref{AFFIL::GermanyMPIK}}$\orcidlink{0000-0002-3778-1432}}, 
  \mbox{W.~Rhode$^{\ref{AFFIL::GermanyUDortmundTU}}$\orcidlink{0000-0003-2636-5000}}, 
  \mbox{M.~Rib\'o$^{\ref{AFFIL::SpainICCUB}}$\orcidlink{0000-0002-9931-4557}}, 
  \mbox{T.~Richtler$^{\ref{AFFIL::ChileUdeConcepcion}}$}, 
  \mbox{F.~Rieger$^{\ref{AFFIL::GermanyMPIK}}$}, 
  \mbox{E.~Roache$^{\ref{AFFIL::USACfAHarvardSmithsonian}}$}, 
  \mbox{G.~Rodriguez~Fernandez$^{\ref{AFFIL::ItalyINFNRomaTorVergata}}$}, 
  \mbox{M.~D.~Rodr{\'\i}guez~Fr{\'\i}as$^{\ref{AFFIL::SpainUAlcala}}$\orcidlink{0000-0002-2550-4462}}, 
  \mbox{J.~J.~Rodr{\'\i}guez-V\'azquez$^{\ref{AFFIL::SpainCIEMAT}}$}, 
  \mbox{P.~Romano$^{\ref{AFFIL::ItalyOBrera}}$\orcidlink{0000-0003-0258-7469}}, 
  \mbox{G.~Romeo$^{\ref{AFFIL::ItalyOCatania}}$\orcidlink{0000-0003-3239-6057}}, 
  \mbox{J.~Rosado$^{\ref{AFFIL::SpainUCMAltasEnergias}}$}, 
  \mbox{G.~Rowell$^{\ref{AFFIL::AustraliaUAdelaide}}$\orcidlink{0000-0002-9516-1581}}, 
  \mbox{B.~Rudak$^{\ref{AFFIL::PolandNicolausCopernicusAstronomicalCenter}}$}, 
  \mbox{A.~J.~Ruiter$^{\ref{AFFIL::AustraliaUNewSouthWalesCanberra}}$\orcidlink{0000-0002-4794-6835}}, 
  \mbox{C.~B.~Rulten$^{\ref{AFFIL::UnitedKingdomUDurham}}$\orcidlink{0000-0001-7483-4348}}, 
  \mbox{F.~Russo$^{\ref{AFFIL::ItalyOASBologna}}$\orcidlink{0000-0002-3476-0839}}, 
  \mbox{I.~Sadeh$^{\ref{AFFIL::GermanyDESY}}$}, 
  \mbox{L.~Saha$^{\ref{AFFIL::USACfAHarvardSmithsonian}}$\orcidlink{0000-0002-3171-5039}}, 
  \mbox{T.~Saito$^{\ref{AFFIL::JapanUTokyoICRR}}$}, 
  \mbox{S.~Sakurai$^{\ref{AFFIL::JapanUTokyoICRR}}$}, 
  \mbox{H.~Salzmann$^{\ref{AFFIL::GermanyIAAT}}$}, 
  \mbox{D.~Sanchez$^{\ref{AFFIL::FranceLAPPUSavoieMontBlanc}}$}, 
  \mbox{M.~S\'anchez-Conde$^{\ref{AFFIL::SpainIFTUAMCSIC}}$\orcidlink{0000-0002-3849-9164}}, 
  \mbox{P.~Sangiorgi$^{\ref{AFFIL::ItalyIASFPalermo}}$\orcidlink{0000-0001-8138-9289}}, 
  \mbox{H.~Sano$^{\ref{AFFIL::JapanUTokyoICRR}}$\orcidlink{0000-0003-2062-5692}}, 
  \mbox{M.~Santander$^{\ref{AFFIL::USAUAlabamaTuscaloosa}}$\orcidlink{0000-0001-7297-8217}}, 
  \mbox{A.~Santangelo$^{\ref{AFFIL::GermanyIAAT}}$}, 
  \mbox{R.~Santos-Lima$^{\ref{AFFIL::BrazilIAGUSaoPaulo}}$\orcidlink{0000-0001-6880-4468}}, 
  \mbox{A.~Sanuy$^{\ref{AFFIL::SpainICCUB}}$}, 
  \mbox{T.~\v{S}ari\'c$^{\ref{AFFIL::CroatiaFESB}}$\orcidlink{0000-0001-8731-8369}}, 
  \mbox{A.~Sarkar$^{\ref{AFFIL::GermanyDESY}}$\orcidlink{0000-0002-7559-4339}}, 
  \mbox{S.~Sarkar$^{\ref{AFFIL::UnitedKingdomUOxford}}$\orcidlink{0000-0002-3542-858X}}, 
  \mbox{F.~G.~Saturni$^{\ref{AFFIL::ItalyORoma}}$\orcidlink{0000-0002-1946-7706}}, 
  \mbox{V.~Savchenko$^{\ref{AFFIL::SwitzerlandEPFLAstroObs}}$\orcidlink{0000-0001-6353-0808}}, 
  \mbox{A.~Scherer$^{\ref{AFFIL::ChileUPontificiaCatolicadeChile}}$}, 
  \mbox{P.~Schipani$^{\ref{AFFIL::ItalyOCapodimonte}}$\orcidlink{0000-0003-0197-589X}}, 
  \mbox{B.~Schleicher$^{\ref{AFFIL::GermanyUWurzburg},\ref{AFFIL::SwitzerlandETHZurich}}$}, 
  \mbox{P.~Schovanek$^{\ref{AFFIL::CzechRepublicFZU}}$}, 
  \mbox{J.~L.~Schubert$^{\ref{AFFIL::GermanyUDortmundTU}}$}, 
  \mbox{F.~Schussler$^{\ref{AFFIL::FranceCEAIRFUDPhP}}$\orcidlink{0000-0003-1500-6571}}, 
  \mbox{U.~Schwanke$^{\ref{AFFIL::GermanyUBerlin}}$\orcidlink{0000-0002-1229-278X}}, 
  \mbox{G.~Schwefer$^{\ref{AFFIL::GermanyMPIK}}$\orcidlink{0000-0002-2050-8413}}, 
  \mbox{S.~Scuderi$^{\ref{AFFIL::ItalyIASFMilano}}$\orcidlink{0000-0002-8637-2109}}, 
  \mbox{M.~Seglar~Arroyo$^{\ref{AFFIL::SpainIFAEBIST}}$\orcidlink{0000-0001-8654-409X}}, 
  \mbox{I.~Seitenzahl$^{\ref{AFFIL::AustraliaUNewSouthWalesCanberra}}$\orcidlink{0000-0002-5044-2988}}, 
  \mbox{O.~Sergijenko$^{\ref{AFFIL::UkraineAstObsofUKyiv},\ref{AFFIL::UkraineObsNASUkraine},\ref{AFFIL::PolandAGHCracowSTC}}$}, 
  \mbox{V.~Sguera$^{\ref{AFFIL::ItalyOASBologna}}$}, 
  \mbox{R.~Y.~Shang$^{\ref{AFFIL::USAUCLA}}$}, 
  \mbox{P.~Sharma$^{\ref{AFFIL::FranceIJCLab}}$}, 
  \mbox{G.~D.~S.~SIDIBE$^{\ref{AFFIL::FranceCEAIRFUDEDIP}}$}, 
  \mbox{L.~Sidoli$^{\ref{AFFIL::ItalyIASFMilano}}$\orcidlink{0000-0001-9705-2883}}, 
  \mbox{H.~Siejkowski$^{\ref{AFFIL::PolandCYFRONETAGH}}$\orcidlink{0000-0003-1673-2145}}, 
  \mbox{C.~Siqueira$^{\ref{AFFIL::BrazilIFSCUSaoPaulo}}$\orcidlink{0000-0001-5684-3849}}, 
  \mbox{P.~Sizun$^{\ref{AFFIL::FranceCEAIRFUDEDIP}}$\orcidlink{0000-0002-8895-3345}}, 
  \mbox{V.~Sliusar$^{\ref{AFFIL::SwitzerlandUGenevaISDC}}$\orcidlink{0000-0002-4387-9372}}, 
  \mbox{A.~Slowikowska$^{\ref{AFFIL::PolandTorunInstituteofAstronomy}}$\orcidlink{0000-0003-4525-3178}}, 
  \mbox{H.~Sol$^{\ref{AFFIL::FranceObservatoiredeParis}}$}, 
  \mbox{A.~Specovius$^{\ref{AFFIL::GermanyUErlangenECAP}}$\orcidlink{0000-0002-1156-4771}}, 
  \mbox{S.~T.~Spencer$^{\ref{AFFIL::GermanyUErlangenECAP},\ref{AFFIL::UnitedKingdomUOxford}}$\orcidlink{0000-0001-5516-1205}}, 
  \mbox{D.~Spiga$^{\ref{AFFIL::ItalyOBrera}}$\orcidlink{0000-0003-1163-7843}}, 
  \mbox{A.~Stamerra$^{\ref{AFFIL::ItalyORoma},\ref{AFFIL::ItalyCTAOBologna}}$\orcidlink{0000-0002-9430-5264}}, 
  \mbox{S.~Stani\v{c}$^{\ref{AFFIL::SloveniaUNovaGoricaCAC}}$\orcidlink{0000-0003-3344-8381}}, 
  \mbox{T.~Starecki$^{\ref{AFFIL::PolandWUTElectronics}}$\orcidlink{0000-0002-4730-6803}}, 
  \mbox{R.~Starling$^{\ref{AFFIL::UnitedKingdomULeicester}}$}, 
  \mbox{C.~Steppa$^{\ref{AFFIL::GermanyUPotsdam}}$}, 
  \mbox{T.~Stolarczyk$^{\ref{AFFIL::FranceCEAIRFUDAp}}$}, 
  \mbox{J.~Stri\v{s}kovi\'c$^{\ref{AFFIL::CroatiaUOsijek}}$}, 
  \mbox{M.~Strzys$^{\ref{AFFIL::JapanUTokyoICRR}}$\orcidlink{0000-0001-5049-1045}}, 
  \mbox{Y.~Suda$^{\ref{AFFIL::JapanUHiroshima}}$\orcidlink{0000-0002-2692-5891}}, 
  \mbox{T.~Suomij\"arvi$^{\ref{AFFIL::FranceIJCLab}}$\orcidlink{0000-0003-1422-258X}}, 
  \mbox{D.~Tak$^{\ref{AFFIL::GermanyDESY}}$\orcidlink{0000-0002-9852-2469}}, 
  \mbox{M.~Takahashi$^{\ref{AFFIL::JapanUNagoyaISEE}}$}, 
  \mbox{R.~Takeishi$^{\ref{AFFIL::JapanUTokyoICRR}}$\orcidlink{0000-0001-6335-5317}}, 
  \mbox{P.-H.~T.~Tam$^{\ref{AFFIL::JapanUTokyoICRR},\ref{AFFIL::ChinaUSunYatsen}}$\orcidlink{0000-0002-1262-7375}}, 
  \mbox{S.~J.~Tanaka$^{\ref{AFFIL::JapanUAoyamaGakuin}}$\orcidlink{0000-0002-8796-1992}}, 
  \mbox{T.~Tanaka$^{\ref{AFFIL::JapanUKonan}}$\orcidlink{0000-0002-4383-0368}}, 
  \mbox{K.~Terauchi$^{\ref{AFFIL::JapanUKyotoPhysicsandAstronomy}}$}, 
  \mbox{V.~Testa$^{\ref{AFFIL::ItalyORoma}}$\orcidlink{0000-0003-1033-1340}}, 
  \mbox{L.~Tibaldo$^{\ref{AFFIL::FranceIRAPUToulouse}}$\orcidlink{0000-0001-7523-570X}}, 
  \mbox{O.~Tibolla$^{\ref{AFFIL::UnitedKingdomUDurham}}$}, 
  \mbox{F.~Torradeflot$^{\ref{AFFIL::SpainPIC},\ref{AFFIL::SpainCIEMAT}}$\orcidlink{0000-0003-1160-1517}}, 
  \mbox{D.~F.~Torres$^{\ref{AFFIL::SpainICECSIC}}$}, 
  \mbox{E.~Torresi$^{\ref{AFFIL::ItalyOASBologna}}$\orcidlink{0000-0002-5201-010X}}, 
  \mbox{N.~Tothill$^{\ref{AFFIL::AustraliaUWesternSydney}}$\orcidlink{0000-0002-9931-5162}}, 
  \mbox{F.~Toussenel$^{\ref{AFFIL::FranceLPNHEUSorbonne}}$}, 
  \mbox{V.~Touzard$^{\ref{AFFIL::FranceIRAPUToulouse}}$}, 
  \mbox{A.~Tramacere$^{\ref{AFFIL::SwitzerlandUGenevaISDC}}$\orcidlink{0000-0002-8186-3793}}, 
  \mbox{P.~Travnicek$^{\ref{AFFIL::CzechRepublicFZU}}$}, 
  \mbox{G.~Tripodo$^{\ref{AFFIL::ItalyUPalermo},\ref{AFFIL::ItalyINFNCatania}}$}, 
  \mbox{S.~Truzzi$^{\ref{AFFIL::ItalyUSienaandINFN}}$}, 
  \mbox{A.~Tsiahina$^{\ref{AFFIL::FranceIRAPUToulouse}}$\orcidlink{0009-0006-6205-8728}}, 
  \mbox{A.~Tutone$^{\ref{AFFIL::ItalyIASFPalermo}}$}, 
  \mbox{M.~Vacula$^{\ref{AFFIL::CzechRepublicUOlomouc},\ref{AFFIL::CzechRepublicFZU}}$\orcidlink{0000-0003-4844-3962}}, 
  \mbox{B.~Vallage$^{\ref{AFFIL::FranceCEAIRFUDPhP}}$\orcidlink{0000-0003-1255-8506}}, 
  \mbox{P.~Vallania$^{\ref{AFFIL::ItalyINFNTorino},\ref{AFFIL::ItalyOTorino}}$}, 
  \mbox{R.~Vall\'es$^{\ref{AFFIL::SpainICECSIC}}$\orcidlink{0000-0001-7701-2163}}, 
  \mbox{C.~van~Eldik$^{\ref{AFFIL::GermanyUErlangenECAP}}$\orcidlink{0000-0001-9669-645X}}, 
  \mbox{J.~van~Scherpenberg$^{\ref{AFFIL::GermanyMPP}}$\orcidlink{0000-0002-6173-867X}}, 
  \mbox{J.~Vandenbroucke$^{\ref{AFFIL::USAUWisconsin}}$}, 
  \mbox{V.~Vassiliev$^{\ref{AFFIL::USAUCLA}}$}, 
  \mbox{P.~Venault$^{\ref{AFFIL::FranceCEAIRFUDEDIP}}$}, 
  \mbox{S.~Ventura$^{\ref{AFFIL::ItalyUSienaandINFN}}$}, 
  \mbox{S.~Vercellone$^{\ref{AFFIL::ItalyOBrera}}$\orcidlink{0000-0003-1163-1396}}, 
  \mbox{G.~Verna$^{\ref{AFFIL::ItalyUSienaandINFN}}$\orcidlink{0000-0001-5916-9028}}, 
  \mbox{A.~Viana$^{\ref{AFFIL::BrazilIFSCUSaoPaulo}}$}, 
  \mbox{N.~Viaux$^{\ref{AFFIL::ChileDepFisUTecnicaFedericoSantaMaria}}$}, 
  \mbox{A.~Vigliano$^{\ref{AFFIL::ItalyUUdineandINFNTrieste}}$}, 
  \mbox{J.~Vignatti$^{\ref{AFFIL::ChileUTecnicaFedericoSantaMaria}}$\orcidlink{0000-0002-1494-9562}}, 
  \mbox{C.~F.~Vigorito$^{\ref{AFFIL::ItalyINFNTorino},\ref{AFFIL::ItalyUTorino}}$\orcidlink{0000-0002-0069-9195}}, 
  \mbox{V.~Vitale$^{\ref{AFFIL::ItalyINFNRomaTorVergata}}$}, 
  \mbox{V.~Vodeb$^{\ref{AFFIL::SloveniaUNovaGoricaCAC}}$}, 
  \mbox{V.~Voisin$^{\ref{AFFIL::FranceLPNHEUSorbonne}}$}, 
  \mbox{S.~Vorobiov$^{\ref{AFFIL::SloveniaUNovaGoricaCAC}}$\orcidlink{0000-0001-8679-3424}}, 
  \mbox{G.~Voutsinas$^{\ref{AFFIL::SwitzerlandUGenevaDPNC}}$}, 
  \mbox{I.~Vovk$^{\ref{AFFIL::JapanUTokyoICRR}}$}, 
  \mbox{V.~Waegebaert$^{\ref{AFFIL::FranceIRAPUToulouse}}$}, 
  \mbox{S.~J.~Wagner$^{\ref{AFFIL::GermanyLSW}}$}, 
  \mbox{R.~Walter$^{\ref{AFFIL::SwitzerlandUGenevaISDC}}$}, 
  \mbox{M.~Ward$^{\ref{AFFIL::UnitedKingdomUDurham}}$}, 
  \mbox{M.~Wechakama$^{\ref{AFFIL::ThailandUKasetsart},\ref{AFFIL::ThailandNARIT}}$\orcidlink{0000-0001-8279-4550}}, 
  \mbox{R.~White$^{\ref{AFFIL::GermanyMPIK}}$}, 
  \mbox{A.~Wierzcholska$^{\ref{AFFIL::PolandIFJ}}$\orcidlink{0000-0003-4472-7204}}, 
  \mbox{M.~Will$^{\ref{AFFIL::GermanyMPP}}$\orcidlink{0000-0002-7504-2083}}, 
  \mbox{D.~A.~Williams$^{\ref{AFFIL::USASCIPP}}$\orcidlink{0000-0003-2740-9714}}, 
  \mbox{F.~Wohlleben$^{\ref{AFFIL::GermanyMPIK}}$\orcidlink{0000-0002-6451-4188}}, 
  \mbox{A.~Wolter$^{\ref{AFFIL::ItalyOBrera}}$\orcidlink{0000-0001-5840-9835}}, 
  \mbox{T.~Yamamoto$^{\ref{AFFIL::JapanUKonan}}$}, 
  \mbox{R.~Yamazaki$^{\ref{AFFIL::JapanUAoyamaGakuin}}$\orcidlink{0000-0002-1251-7889}}, 
  \mbox{L.~Yang$^{\ref{AFFIL::SouthAfricaUJohannesburg},\ref{AFFIL::ChinaUSunYatsen}}$}, 
  \mbox{T.~Yoshida$^{\ref{AFFIL::JapanUIbaraki}}$}, 
  \mbox{T.~Yoshikoshi$^{\ref{AFFIL::JapanUTokyoICRR}}$\orcidlink{0000-0002-6045-9839}}, 
  \mbox{M.~Zacharias$^{\ref{AFFIL::GermanyLSW},\ref{AFFIL::SouthAfricaNWU}}$\orcidlink{0000-0001-5801-3945}}, 
  \mbox{R.~Zanmar~Sanchez$^{\ref{AFFIL::ItalyOCatania}}$\orcidlink{0000-0002-6997-0887}}, 
  \mbox{D.~Zavrtanik$^{\ref{AFFIL::SloveniaUNovaGoricaCAC}}$\orcidlink{0000-0002-4596-1521}}, 
  \mbox{M.~Zavrtanik$^{\ref{AFFIL::SloveniaUNovaGoricaCAC}}$}, 
  \mbox{A.~A.~Zdziarski$^{\ref{AFFIL::PolandNicolausCopernicusAstronomicalCenter}}$}, 
  \mbox{A.~Zech$^{\ref{AFFIL::FranceObservatoiredeParis}}$\orcidlink{0000-0002-4388-5625}}, 
  \mbox{V.~I.~Zhdanov$^{\ref{AFFIL::UkraineAstObsofUKyiv}}$\orcidlink{0000-0003-3690-483X}}, 
  \mbox{K.~Zi\k{e}tara$^{\ref{AFFIL::PolandUJagiellonian}}$}, 
  \mbox{M.~\v{Z}ivec$^{\ref{AFFIL::SloveniaUNovaGoricaCAC}}$}, 
  \mbox{J.~Zuriaga-Puig$^{\ref{AFFIL::SpainIFTUAMCSIC}}$\orcidlink{0000-0003-0652-6700}}
\twocolumn
\section*{Affiliations}
\begin{enumerate}[label=$^{\arabic*}$,ref=\arabic*,leftmargin=1.5em,labelsep=0.25em,labelwidth=1.25em]
\item Department of Physics, Tokai University, 4-1-1, Kita-Kaname, Hiratsuka, Kanagawa 259-1292, Japan\label{AFFIL::JapanUTokai}
\item Institute for Cosmic Ray Research, University of Tokyo, 5-1-5, Kashiwa-no-ha, Kashiwa, Chiba 277-8582, Japan\label{AFFIL::JapanUTokyoICRR}
\item University of Alabama, Tuscaloosa, Department of Physics and Astronomy, Gallalee Hall, Box 870324 Tuscaloosa, AL 35487-0324, USA\label{AFFIL::USAUAlabamaTuscaloosa}
\item Universit\'e C\^ote d'Azur, Observatoire de la C\^ote d'Azur, CNRS, Laboratoire Lagrange, France\label{AFFIL::FranceOCotedAzur}
\item Laboratoire Leprince-Ringuet, CNRS/IN2P3, \'Ecole polytechnique, Institut Polytechnique de Paris, 91120 Palaiseau, France\label{AFFIL::FranceLLREcolePolytechnique}
\item Departament de F{\'\i}sica Qu\`antica i Astrof{\'\i}sica, Institut de Ci\`encies del Cosmos, Universitat de Barcelona, IEEC-UB, Mart{\'\i} i Franqu\`es, 1, 08028, Barcelona, Spain\label{AFFIL::SpainICCUB}
\item Instituto de Astrof{\'\i}sica de Andaluc{\'\i}a-CSIC, Glorieta de la Astronom{\'\i}a s/n, 18008, Granada, Spain\label{AFFIL::SpainIAACSIC}
\item Pontificia Universidad Cat\'olica de Chile, Av. Libertador Bernardo O'Higgins 340, Santiago, Chile\label{AFFIL::ChileUPontificiaCatolicadeChile}
\item IPARCOS-UCM, Instituto de F{\'\i}sica de Part{\'\i}culas y del Cosmos, and EMFTEL Department, Universidad Complutense de Madrid, E-28040 Madrid, Spain\label{AFFIL::SpainUCMAltasEnergias}
\item Instituto de F{\'\i}sica Te\'orica UAM/CSIC and Departamento de F{\'\i}sica Te\'orica, Universidad Aut\'onoma de Madrid, c/ Nicol\'as Cabrera 13-15, Campus de Cantoblanco UAM, 28049 Madrid, Spain\label{AFFIL::SpainIFTUAMCSIC}
\item LUTH, GEPI and LERMA, Observatoire de Paris, Universit\'e PSL, Universit\'e Paris Cit\'e, CNRS, 5 place Jules Janssen, 92190, Meudon, France\label{AFFIL::FranceObservatoiredeParis}
\item INAF - Osservatorio Astrofisico di Arcetri, Largo E. Fermi, 5 - 50125 Firenze, Italy\label{AFFIL::ItalyOArcetri}
\item INAF - Osservatorio Astronomico di Roma, Via di Frascati 33, 00040, Monteporzio Catone, Italy\label{AFFIL::ItalyORoma}
\item T\"UB\.ITAK Research Institute for Fundamental Sciences, 41470 Gebze, Kocaeli, Turkey\label{AFFIL::TurkeyTubitak}
\item INFN Sezione di Napoli, Via Cintia, ed. G, 80126 Napoli, Italy\label{AFFIL::ItalyINFNNapoli}
\item INFN Sezione di Padova, Via Marzolo 8, 35131 Padova, Italy\label{AFFIL::ItalyINFNPadova}
\item Laboratoire Univers et Particules de Montpellier, Universit\'e de Montpellier, CNRS/IN2P3, CC 72, Place Eug\`ene Bataillon, F-34095 Montpellier Cedex 5, France\label{AFFIL::FranceLUPMUMontpellier}
\item Kapteyn Astronomical Institute, University of Groningen, Landleven 12, 9747 AD, Groningen, The Netherlands\label{AFFIL::NetherlandsUGroningen}
\item Instituto de F{\'\i}sica de S\~ao Carlos, Universidade de S\~ao Paulo, Av. Trabalhador S\~ao-carlense, 400 - CEP 13566-590, S\~ao Carlos, SP, Brazil\label{AFFIL::BrazilIFSCUSaoPaulo}
\item Astroparticle Physics, Department of Physics, TU Dortmund University, Otto-Hahn-Str. 4a, 44227 Dortmund, Germany\label{AFFIL::GermanyUDortmundTU}
\item Department of Physics, Chemistry \& Material Science, University of Namibia, Private Bag 13301, Windhoek, Namibia\label{AFFIL::NamibiaUNamibia}
\item Centre for Space Research, North-West University, Potchefstroom, 2520, South Africa\label{AFFIL::SouthAfricaNWU}
\item School of Physics and Astronomy, Monash University, Melbourne, Victoria 3800, Australia\label{AFFIL::AustraliaUMonash}
\item Department of Astronomy, University of Geneva, Chemin d'Ecogia 16, CH-1290 Versoix, Switzerland\label{AFFIL::SwitzerlandUGenevaISDC}
\item Faculty of Science and Technology, Universidad del Azuay, Cuenca, Ecuador.\label{AFFIL::EcuadorUAzuay}
\item Deutsches Elektronen-Synchrotron, Platanenallee 6, 15738 Zeuthen, Germany\label{AFFIL::GermanyDESY}
\item Centro Brasileiro de Pesquisas F{\'\i}sicas, Rua Xavier Sigaud 150, RJ 22290-180, Rio de Janeiro, Brazil\label{AFFIL::BrazilCBPF}
\item Instituto de Astronomia, Geof{\'\i}sica e Ci\^encias Atmosf\'ericas - Universidade de S\~ao Paulo, Cidade Universit\'aria, R. do Mat\~ao, 1226, CEP 05508-090, S\~ao Paulo, SP, Brazil\label{AFFIL::BrazilIAGUSaoPaulo}
\item INFN Sezione di Padova and Universit\`a degli Studi di Padova, Via Marzolo 8, 35131 Padova, Italy\label{AFFIL::ItalyUPadovaandINFN}
\item Institut f\"ur Physik \& Astronomie, Universit\"at Potsdam, Karl-Liebknecht-Strasse 24/25, 14476 Potsdam, Germany\label{AFFIL::GermanyUPotsdam}
\item University of the Witwatersrand, 1 Jan Smuts Avenue, Braamfontein, 2000 Johannesburg, South Africa\label{AFFIL::SouthAfricaUWitwatersrand}
\item Institut f\"ur Theoretische Physik, Lehrstuhl IV: Plasma-Astroteilchenphysik, Ruhr-Universit\"at Bochum, Universit\"atsstra{\ss}e 150, 44801 Bochum, Germany\label{AFFIL::GermanyUBochum}
\item INFN Sezione di Roma Tor Vergata, Via della Ricerca Scientifica 1, 00133 Rome, Italy\label{AFFIL::ItalyINFNRomaTorVergata}
\item Center for Astrophysics | Harvard \& Smithsonian, 60 Garden St, Cambridge, MA 02138, USA\label{AFFIL::USACfAHarvardSmithsonian}
\item CIEMAT, Avda. Complutense 40, 28040 Madrid, Spain\label{AFFIL::SpainCIEMAT}
\item Max-Planck-Institut f\"ur Kernphysik, Saupfercheckweg 1, 69117 Heidelberg, Germany\label{AFFIL::GermanyMPIK}
\item Max-Planck-Institut f\"ur Physik, F\"ohringer Ring 6, 80805 M\"unchen, Germany\label{AFFIL::GermanyMPP}
\item Pidstryhach Institute for Applied Problems in Mechanics and Mathematics NASU, 3B Naukova Street, Lviv, 79060, Ukraine\label{AFFIL::UkraineIAPMMLviv}
\item Univ. Savoie Mont Blanc, CNRS, Laboratoire d'Annecy de Physique des Particules - IN2P3, 74000 Annecy, France\label{AFFIL::FranceLAPPUSavoieMontBlanc}
\item Center for Astrophysics and Cosmology (CAC), University of Nova Gorica, Nova Gorica, Slovenia\label{AFFIL::SloveniaUNovaGoricaCAC}
\item Institut f\"ur Astronomie und Astrophysik, Universit\"at T\"ubingen, Sand 1, 72076 T\"ubingen, Germany\label{AFFIL::GermanyIAAT}
\item ETH Z\"urich, Institute for Particle Physics and Astrophysics, Otto-Stern-Weg 5, 8093 Z\"urich, Switzerland\label{AFFIL::SwitzerlandETHZurich}
\item Politecnico di Bari, via Orabona 4, 70124 Bari, Italy\label{AFFIL::ItalyPolitecnicoBari}
\item INFN Sezione di Bari, via Orabona 4, 70126 Bari, Italy\label{AFFIL::ItalyINFNBari}
\item Institut de Fisica d'Altes Energies (IFAE), The Barcelona Institute of Science and Technology, Campus UAB, 08193 Bellaterra (Barcelona), Spain\label{AFFIL::SpainIFAEBIST}
\item FZU - Institute of Physics of the Czech Academy of Sciences, Na Slovance 1999/2, 182 21 Praha 8, Czech Republic\label{AFFIL::CzechRepublicFZU}
\item Sorbonne Universit\'e, CNRS/IN2P3, Laboratoire de Physique Nucl\'eaire et de Hautes Energies, LPNHE, 4 place Jussieu, 75005 Paris, France\label{AFFIL::FranceLPNHEUSorbonne}
\item INAF - Osservatorio Astronomico di Brera, Via Brera 28, 20121 Milano, Italy\label{AFFIL::ItalyOBrera}
\item INFN Sezione di Pisa, Edificio C {\textendash} Polo Fibonacci, Largo Bruno Pontecorvo 3, 56127 Pisa\label{AFFIL::ItalyINFNPisa}
\item University of Zagreb, Faculty of electrical engineering and computing, Unska 3, 10000 Zagreb, Croatia\label{AFFIL::CroatiaUZagreb}
\item IRFU, CEA, Universit\'e Paris-Saclay, B\^at 141, 91191 Gif-sur-Yvette, France\label{AFFIL::FranceCEAIRFUDPhP}
\item School of Physics, Chemistry and Earth Sciences, University of Adelaide, Adelaide SA 5005, Australia\label{AFFIL::AustraliaUAdelaide}
\item INAF - Osservatorio di Astrofisica e Scienza dello spazio di Bologna, Via Piero Gobetti 93/3, 40129  Bologna, Italy\label{AFFIL::ItalyOASBologna}
\item Dublin Institute for Advanced Studies, 31 Fitzwilliam Place, Dublin 2, Ireland\label{AFFIL::IrelandDIAS}
\item Centre for Advanced Instrumentation, Department of Physics, Durham University, South Road, Durham, DH1 3LE, United Kingdom\label{AFFIL::UnitedKingdomUDurham}
\item INFN Sezione di Trieste and Universit\`a degli Studi di Udine, Via delle Scienze 208, 33100 Udine, Italy\label{AFFIL::ItalyUUdineandINFNTrieste}
\item University of Geneva - D\'epartement de physique nucl\'eaire et corpusculaire, 24 rue du G\'en\'eral-Dufour, 1211 Gen\`eve 4, Switzerland\label{AFFIL::SwitzerlandUGenevaDPNC}
\item Armagh Observatory and Planetarium, College Hill, Armagh BT61 9DG, United Kingdom\label{AFFIL::UnitedKingdomArmaghObservatoryandPlanetarium}
\item School of Physics, University of New South Wales, Sydney NSW 2052, Australia\label{AFFIL::AustraliaUNewSouthWales}
\item Universit\'e Paris-Saclay, Universit\'e Paris Cit\'e, CEA, CNRS, AIM, F-91191 Gif-sur-Yvette Cedex, France\label{AFFIL::FranceCEAIRFUDAp}
\item Cherenkov Telescope Array Observatory, Saupfercheckweg 1, 69117 Heidelberg, Germany\label{AFFIL::GermanyCTAOHeidelberg}
\item Unitat de F{\'\i}sica de les Radiacions, Departament de F{\'\i}sica, and CERES-IEEC, Universitat Aut\`onoma de Barcelona, Edifici C3, Campus UAB, 08193 Bellaterra, Spain\label{AFFIL::SpainUABandCERESIEEC}
\item Department of Physics, Faculty of Science, Kasetsart University, 50 Ngam Wong Wan Rd., Lat Yao, Chatuchak, Bangkok, 10900, Thailand\label{AFFIL::ThailandUKasetsart}
\item National Astronomical Research Institute of Thailand, 191 Huay Kaew Rd., Suthep, Muang, Chiang Mai, 50200, Thailand\label{AFFIL::ThailandNARIT}
\item INAF - Istituto di Astrofisica Spaziale e Fisica Cosmica di Palermo, Via U. La Malfa 153, 90146 Palermo, Italy\label{AFFIL::ItalyIASFPalermo}
\item Universidade Cruzeiro do Sul, N\'ucleo de Astrof{\'\i}sica Te\'orica (NAT/UCS), Rua Galv\~ao Bueno 8687, Bloco B, sala 16, Libertade 01506-000 - S\~ao Paulo, Brazil\label{AFFIL::BrazilUCruzeirodoSul}
\item Lund Observatory, Lund University, Box 43, SE-22100 Lund, Sweden\label{AFFIL::SwedenLundObservatory}
\item Aix Marseille Univ, CNRS/IN2P3, CPPM, Marseille, France\label{AFFIL::FranceCPPMUAixMarseille}
\item INAF - Osservatorio Astronomico di Capodimonte, Via Salita Moiariello 16, 80131 Napoli, Italy\label{AFFIL::ItalyOCapodimonte}
\item Universidad de Alcal\'a - Space \& Astroparticle group, Facultad de Ciencias, Campus Universitario Ctra. Madrid-Barcelona, Km. 33.600 28871 Alcal\'a de Henares (Madrid), Spain\label{AFFIL::SpainUAlcala}
\item Escola de Engenharia de Lorena, Universidade de S\~ao Paulo, \'Area I - Estrada Municipal do Campinho, s/n{\textdegree}, CEP 12602-810, Pte. Nova, Lorena, Brazil\label{AFFIL::BrazilEELUSaoPaulo}
\item INFN Sezione di Bari and Universit\`a degli Studi di Bari, via Orabona 4, 70124 Bari, Italy\label{AFFIL::ItalyUandINFNBari}
\item Universit\'e Paris Cit\'e, CNRS, Astroparticule et Cosmologie, F-75013 Paris, France\label{AFFIL::FranceAPCUParisCite}
\item Dublin City University, Glasnevin, Dublin 9, Ireland\label{AFFIL::IrelandDCU}
\item INFN Sezione di Torino, Via P. Giuria 1, 10125 Torino, Italy\label{AFFIL::ItalyINFNTorino}
\item Dipartimento di Fisica - Universit\`a degli Studi di Torino, Via Pietro Giuria 1 - 10125 Torino, Italy\label{AFFIL::ItalyUTorino}
\item Universidade Federal Do Paran\'a - Setor Palotina, Departamento de Engenharias e Exatas, Rua Pioneiro, 2153, Jardim Dallas, CEP: 85950-000 Palotina, Paran\'a, Brazil\label{AFFIL::BrazilUFPR}
\item INAF - Osservatorio Astrofisico di Catania, Via S. Sofia, 78, 95123 Catania, Italy\label{AFFIL::ItalyOCatania}
\item Universidad de Valpara{\'\i}so, Blanco 951, Valparaiso, Chile\label{AFFIL::ChileUdeValparaiso}
\item University of Wisconsin, Madison, 500 Lincoln Drive, Madison, WI, 53706, USA\label{AFFIL::USAUWisconsin}
\item Department of Physics and Technology, University of Bergen, Museplass 1, 5007 Bergen, Norway\label{AFFIL::NorwayUBergen}
\item INAF - Istituto di Radioastronomia, Via Gobetti 101, 40129 Bologna, Italy\label{AFFIL::ItalyRadioastronomiaINAF}
\item INAF - Istituto Nazionale di Astrofisica, Viale del Parco Mellini 84, 00136 Rome, Italy\label{AFFIL::ItalyINAF}
\item IRFU/DEDIP, CEA, Universit\'e Paris-Saclay, Bat 141, 91191 Gif-sur-Yvette, France\label{AFFIL::FranceCEAIRFUDEDIP}
\item Universit\'a degli Studi di Napoli {\textquotedblleft}Federico II{\textquotedblright} - Dipartimento di Fisica {\textquotedblleft}E. Pancini{\textquotedblright}, Complesso universitario di Monte Sant'Angelo, Via Cintia - 80126 Napoli, Italy\label{AFFIL::ItalyUNapoli}
\item CCTVal, Universidad T\'ecnica Federico Santa Mar{\'\i}a, Avenida Espa\~na 1680, Valpara{\'\i}so, Chile\label{AFFIL::ChileUTecnicaFedericoSantaMaria}
\item Institute for Theoretical Physics and Astrophysics, Universit\"at W\"urzburg, Campus Hubland Nord, Emil-Fischer-Str. 31, 97074 W\"urzburg, Germany\label{AFFIL::GermanyUWurzburg}
\item Friedrich-Alexander-Universit\"at Erlangen-N\"urnberg, Erlangen Centre for Astroparticle Physics, Nikolaus-Fiebiger-Str. 2, 91058 Erlangen, Germany\label{AFFIL::GermanyUErlangenECAP}
\item Universit\'e Paris-Saclay, CNRS/IN2P3, IJCLab, 91405 Orsay, France\label{AFFIL::FranceIJCLab}
\item Department of Astronomy and Astrophysics, University of Chicago, 5640 S Ellis Ave, Chicago, Illinois, 60637, USA\label{AFFIL::USAUChicagoDAA}
\item LAPTh, CNRS, USMB, F-74940 Annecy, France\label{AFFIL::FranceLAPTh}
\item Santa Cruz Institute for Particle Physics and Department of Physics, University of California, Santa Cruz, 1156 High Street, Santa Cruz, CA 95064, USA\label{AFFIL::USASCIPP}
\item University School for Advanced Studies IUSS Pavia, Palazzo del Broletto, Piazza della Vittoria 15, 27100 Pavia, Italy\label{AFFIL::ItalyIUSSPaviaINAF}
\item INAF - Istituto di Astrofisica Spaziale e Fisica Cosmica di Milano, Via A. Corti 12, 20133 Milano, Italy\label{AFFIL::ItalyIASFMilano}
\item Escola de Artes, Ci\^encias e Humanidades, Universidade de S\~ao Paulo, Rua Arlindo Bettio, CEP 03828-000, 1000 S\~ao Paulo, Brazil\label{AFFIL::BrazilEACHUSaoPaulo}
\item Astronomical Observatory of Taras Shevchenko National University of Kyiv, 3 Observatorna Street, Kyiv, 04053, Ukraine\label{AFFIL::UkraineAstObsofUKyiv}
\item The University of Manitoba, Dept of Physics and Astronomy, Winnipeg, Manitoba R3T 2N2, Canada\label{AFFIL::CanadaUManitoba}
\item RIKEN, Institute of Physical and Chemical Research, 2-1 Hirosawa, Wako, Saitama, 351-0198, Japan\label{AFFIL::JapanRIKEN}
\item INFN Sezione di Roma La Sapienza, P.le Aldo Moro, 2 - 00185 Roma, Italy\label{AFFIL::ItalyINFNRomaLaSapienza}
\item INFN Sezione di Perugia and Universit\`a degli Studi di Perugia, Via A. Pascoli, 06123 Perugia, Italy\label{AFFIL::ItalyUPerugiaandINFN}
\item INAF - Istituto di Astrofisica e Planetologia Spaziali (IAPS), Via del Fosso del Cavaliere 100, 00133 Roma, Italy\label{AFFIL::ItalyIAPS}
\item Department of Physics, Nagoya University, Chikusa-ku, Nagoya, 464-8602, Japan\label{AFFIL::JapanUNagoya}
\item Alikhanyan National Science Laboratory, Yerevan Physics Institute, 2 Alikhanyan Brothers St., 0036, Yerevan, Armenia\label{AFFIL::ArmeniaNSLAlikhanyan}
\item INFN Sezione di Catania, Via S. Sofia 64, 95123 Catania, Italy\label{AFFIL::ItalyINFNCatania}
\item Universit\'e Paris Cit\'e, CNRS, CEA, Astroparticule et Cosmologie, F-75013 Paris, France\label{AFFIL::FranceAPCUParisCiteCEAaffiliatedpersonnel}
\item Universidad Andres Bello, Rep\'ublica 252, Santiago, Chile\label{AFFIL::ChileUAndresBello}
\item Universidad Nacional Aut\'onoma de M\'exico, Delegaci\'on Coyoac\'an, 04510 Ciudad de M\'exico, Mexico\label{AFFIL::MexicoUNAMMexico}
\item N\'ucleo de Astrof{\'\i}sica e Cosmologia (Cosmo-ufes) \& Departamento de F{\'\i}sica, Universidade Federal do Esp{\'\i}rito Santo (UFES), Av. Fernando Ferrari, 514. 29065-910. Vit\'oria-ES, Brazil\label{AFFIL::BrazilUFES}
\item Astrophysics Research Center of the Open University (ARCO), The Open University of Israel, P.O. Box 808, Ra{\textquoteright}anana 4353701, Israel\label{AFFIL::IsraelOpenUniversityofIsrael}
\item Department of Physics, The George Washington University, Washington, DC 20052, USA\label{AFFIL::USAGWUWashingtonDC}
\item University of Liverpool, Oliver Lodge Laboratory, Liverpool L69 7ZE, United Kingdom\label{AFFIL::UnitedKingdomULiverpool}
\item King's College London, Strand, London, WC2R 2LS, United Kingdom\label{AFFIL::UnitedKingdomKingsCollege}
\item Department of Physics, Yamagata University, Yamagata, Yamagata 990-8560, Japan\label{AFFIL::JapanUYamagata}
\item Learning and Education Development Center, Yamanashi-Gakuin University, Kofu, Yamanashi 400-8575, Japan\label{AFFIL::JapanUYamanashiGakuin}
\item IRAP, Universit\'e de Toulouse, CNRS, CNES, UPS, 9 avenue Colonel Roche, 31028 Toulouse, Cedex 4, France\label{AFFIL::FranceIRAPUToulouse}
\item Universit\"at Innsbruck, Institut f\"ur Astro- und Teilchenphysik, Technikerstr. 25/8, 6020 Innsbruck, Austria\label{AFFIL::AustriaUInnsbruck}
\item Palack\'y University Olomouc, Faculty of Science, Joint Laboratory of Optics of Palack\'y University and Institute of Physics of the Czech Academy of Sciences, 17. listopadu 1192/12, 779 00 Olomouc, Czech Republic\label{AFFIL::CzechRepublicUOlomouc}
\item Finnish Centre for Astronomy with ESO, University of Turku, Finland, FI-20014 University of Turku, Finland\label{AFFIL::FinlandUTurku}
\item Josip Juraj Strossmayer University of Osijek, Trg Ljudevita Gaja 6, 31000 Osijek, Croatia\label{AFFIL::CroatiaUOsijek}
\item Gran Sasso Science Institute (GSSI), Viale Francesco Crispi 7, 67100 L{\textquoteright}Aquila, Italy and INFN-Laboratori Nazionali del Gran Sasso (LNGS), via G. Acitelli 22, 67100 Assergi (AQ), Italy\label{AFFIL::ItalyGSSIandINFNAquila}
\item Dipartimento di Scienze Fisiche e Chimiche, Universit\`a degli Studi dell'Aquila and GSGC-LNGS-INFN, Via Vetoio 1, L'Aquila, 67100, Italy\label{AFFIL::ItalyUandINFNAquila}
\item Faculty of Physics and Applied Computer Science,  University of L\'od\'z, ul. Pomorska 149-153, 90-236 L\'od\'z, Poland\label{AFFIL::PolandULodz}
\item Astronomical Observatory, Jagiellonian University, ul. Orla 171, 30-244 Cracow, Poland\label{AFFIL::PolandUJagiellonian}
\item Landessternwarte, Zentrum f\"ur Astronomie  der Universit\"at Heidelberg, K\"onigstuhl 12, 69117 Heidelberg, Germany\label{AFFIL::GermanyLSW}
\item Univ. Grenoble Alpes, CNRS, IPAG, 414 rue de la Piscine, Domaine Universitaire, 38041 Grenoble Cedex 9, France\label{AFFIL::FranceIPAGUGrenobleAlpes}
\item Astronomical Institute of the Czech Academy of Sciences, Bocni II 1401 - 14100 Prague, Czech Republic\label{AFFIL::CzechRepublicASU}
\item Department of Physics and Astronomy, University of Utah, Salt Lake City, UT 84112-0830, USA\label{AFFIL::USAUUtah}
\item Nicolaus Copernicus Astronomical Center, Polish Academy of Sciences, ul. Bartycka 18, 00-716 Warsaw, Poland\label{AFFIL::PolandNicolausCopernicusAstronomicalCenter}
\item Institute of Particle and Nuclear Studies,  KEK (High Energy Accelerator Research Organization), 1-1 Oho, Tsukuba, 305-0801, Japan\label{AFFIL::JapanKEK}
\item School of Physics and Astronomy, University of Leicester, Leicester, LE1 7RH, United Kingdom\label{AFFIL::UnitedKingdomULeicester}
\item Western Sydney University, Locked Bag 1797, Penrith, NSW 2751, Australia\label{AFFIL::AustraliaUWesternSydney}
\item Universit\'e Bordeaux, CNRS, LP2I Bordeaux, UMR 5797, 19 Chemin du Solarium, F-33170 Gradignan, France\label{AFFIL::FranceLP2IUBordeaux}
\item INFN Sezione di Trieste and Universit\`a degli Studi di Trieste, Via Valerio 2 I, 34127 Trieste, Italy\label{AFFIL::ItalyUandINFNTrieste}
\item Instituto de Astrof{\'\i}sica de Canarias and Departamento de Astrof{\'\i}sica, Universidad de La Laguna, La Laguna, Tenerife, Spain\label{AFFIL::SpainIAC}
\item Escuela Polit\'ecnica Superior de Ja\'en, Universidad de Ja\'en, Campus Las Lagunillas s/n, Edif. A3, 23071 Ja\'en, Spain\label{AFFIL::SpainUJaen}
\item Anton Pannekoek Institute/GRAPPA, University of Amsterdam, Science Park 904 1098 XH Amsterdam, The Netherlands\label{AFFIL::NetherlandsUAmsterdam}
\item Saha Institute of Nuclear Physics, A CI of Homi Bhabha National Institute, Kolkata 700064, West Bengal, India\label{AFFIL::IndiaSahaInstitute}
\item Universit\`a degli studi di Catania, Dipartimento di Fisica e Astronomia {\textquotedblleft}Ettore Majorana{\textquotedblright}, Via S. Sofia 64, 95123 Catania, Italy\label{AFFIL::ItalyUCatania}
\item Dipartimento di Fisica e Chimica {\textquotedblleft}E. Segr\`e{\textquotedblright}, Universit\`a degli Studi di Palermo, Via Archirafi 36, 90123, Palermo, Italy\label{AFFIL::ItalyUPalermo}
\item UCM-ELEC group, EMFTEL Department, University Complutense of Madrid, 28040 Madrid, Spain\label{AFFIL::SpainUCMElectronica}
\item Departamento de Ingenier{\'\i}a El\'ectrica, Universidad Pontificia de Comillas - ICAI, 28015 Madrid\label{AFFIL::SpainUPCMadrid}
\item Universidad de Chile, Av. Libertador Bernardo O'Higgins 1058, Santiago, Chile\label{AFFIL::ChileUdeChile}
\item Institute of Space Sciences (ICE, CSIC), and Institut d'Estudis Espacials de Catalunya (IEEC), and Instituci\'o Catalana de Recerca I Estudis Avan\c{c}ats (ICREA), Campus UAB, Carrer de Can Magrans, s/n 08193 Cerdanyola del Vall\'es, Spain\label{AFFIL::SpainICECSIC}
\item The Henryk Niewodnicza\'nski Institute of Nuclear Physics, Polish Academy of Sciences, ul. Radzikowskiego 152, 31-342 Cracow, Poland\label{AFFIL::PolandIFJ}
\item IPARCOS Institute, Faculty of Physics (UCM), 28040 Madrid, Spain\label{AFFIL::SpainIPARCOSInstitute}
\item Department of Physics, Konan University, Kobe, Hyogo, 658-8501, Japan\label{AFFIL::JapanUKonan}
\item Hiroshima Astrophysical Science Center, Hiroshima University, Higashi-Hiroshima, Hiroshima 739-8526, Japan\label{AFFIL::JapanHASC}
\item Department of Physics, Columbia University, 538 West 120th Street, New York, NY 10027, USA\label{AFFIL::USABarnardCollegeColumbiaUniversity}
\item School of Allied Health Sciences, Kitasato University, Sagamihara, Kanagawa 228-8555, Japan\label{AFFIL::JapanUKitasato}
\item Kavli Institute for Particle Astrophysics and Cosmology, Stanford University, Stanford, CA 94305, USA\label{AFFIL::USAStanford}
\item University of Bia{\l}ystok, Faculty of Physics, ul. K. Cio{\l}kowskiego 1L, 15-245 Bia{\l}ystok, Poland\label{AFFIL::PolandUBiaystok}
\item Charles University, Institute of Particle \& Nuclear Physics, V Hole\v{s}ovi\v{c}k\'ach 2, 180 00 Prague 8, Czech Republic\label{AFFIL::CzechRepublicUPrague}
\item Astronomical Observatory of Ivan Franko National University of Lviv, 8 Kyryla i Mephodia Street, Lviv, 79005, Ukraine\label{AFFIL::UkraineAstObsofULviv}
\item Institute for Space{\textemdash}Earth Environmental Research, Nagoya University, Furo-cho, Chikusa-ku, Nagoya 464-8601, Japan\label{AFFIL::JapanUNagoyaISEE}
\item Kobayashi{\textemdash}Maskawa Institute for the Origin of Particles and the Universe, Nagoya University, Furo-cho, Chikusa-ku, Nagoya 464-8602, Japan\label{AFFIL::JapanUNagoyaKMI}
\item INAF - Osservatorio Astronomico di Palermo {\textquotedblleft}G.S. Vaiana{\textquotedblright}, Piazza del Parlamento 1, 90134 Palermo, Italy\label{AFFIL::ItalyOPalermo}
\item Department of Physics and Astronomy, University of California, Los Angeles, CA 90095, USA\label{AFFIL::USAUCLA}
\item Graduate School of Technology, Industrial and Social Sciences, Tokushima University, Tokushima 770-8506, Japan\label{AFFIL::JapanUTokushima}
\item School of Physics \& Center for Relativistic Astrophysics, Georgia Institute of Technology, 837 State Street, Atlanta, Georgia, 30332-0430, USA\label{AFFIL::USAGeorgiaTech}
\item University of Pisa, Largo B. Pontecorvo 3, 56127 Pisa, Italy \label{AFFIL::ItalyUPisa}
\item University of Rijeka, Faculty of Physics, Radmile Matejcic 2, 51000 Rijeka, Croatia\label{AFFIL::CroatiaURijeka}
\item Rudjer Boskovic Institute, Bijenicka 54, 10 000 Zagreb, Croatia\label{AFFIL::CroatiaIRB}
\item INAF - Osservatorio Astronomico di Padova, Vicolo dell'Osservatorio 5, 35122 Padova, Italy\label{AFFIL::ItalyOPadova}
\item INAF - Osservatorio Astronomico di Padova and INFN Sezione di Trieste, gr. coll. Udine, Via delle Scienze 208 I-33100 Udine, Italy\label{AFFIL::ItalyOandINFNTrieste}
\item INFN and Universit\`a degli Studi di Siena, Dipartimento di Scienze Fisiche, della Terra e dell'Ambiente (DSFTA), Sezione di Fisica, Via Roma 56, 53100 Siena, Italy\label{AFFIL::ItalyUSienaandINFN}
\item Centre for Astro-Particle Physics (CAPP) and Department of Physics, University of Johannesburg, PO Box 524, Auckland Park 2006, South Africa\label{AFFIL::SouthAfricaUJohannesburg}
\item University of Oxford, Department of Physics, Clarendon Laboratory, Parks Road, Oxford, OX1 3PU, United Kingdom\label{AFFIL::UnitedKingdomUOxford}
\item Departamento de F{\'\i}sica, Facultad de Ciencias B\'asicas, Universidad Metropolitana de Ciencias de la Educaci\'on, Avenida Jos\'e Pedro Alessandri 774, \~Nu\~noa, Santiago, Chile\label{AFFIL::ChileUMCE}
\item Departamento de Astronom{\'\i}a, Universidad de Concepci\'on, Barrio Universitario S/N, Concepci\'on, Chile\label{AFFIL::ChileUdeConcepcion}
\item University of New South Wales, School of Science, Australian Defence Force Academy, Canberra, ACT 2600, Australia \label{AFFIL::AustraliaUNewSouthWalesCanberra}
\item University of Split  - FESB, R. Boskovica 32, 21 000 Split, Croatia\label{AFFIL::CroatiaFESB}
\item EPFL Laboratoire d{\textquoteright}astrophysique, Observatoire de Sauverny, CH-1290 Versoix, Switzerland\label{AFFIL::SwitzerlandEPFLAstroObs}
\item Department of Physics, Humboldt University Berlin, Newtonstr. 15, 12489 Berlin, Germany\label{AFFIL::GermanyUBerlin}
\item Main Astronomical Observatory of the National Academy of Sciences of Ukraine, Zabolotnoho str., 27, 03143, Kyiv, Ukraine\label{AFFIL::UkraineObsNASUkraine}
\item Space Technology Centre, AGH University of Science and Technology, Aleja Mickiewicza, 30, 30-059, Krak\'ow, Poland\label{AFFIL::PolandAGHCracowSTC}
\item Academic Computer Centre CYFRONET AGH, ul. Nawojki 11, 30-950, Krak\'ow, Poland\label{AFFIL::PolandCYFRONETAGH}
\item Institute of Astronomy, Faculty of Physics, Astronomy and Informatics, Nicolaus Copernicus University in Toru\'n, ul. Grudzi\k{a}dzka 5, 87-100 Toru\'n, Poland\label{AFFIL::PolandTorunInstituteofAstronomy}
\item Cherenkov Telescope Array Observatory gGmbH, Via Gobetti, Bologna, Italy\label{AFFIL::ItalyCTAOBologna}
\item Warsaw University of Technology, Faculty of Electronics and Information Technology, Institute of Electronic Systems, Nowowiejska 15/19, 00-665 Warsaw, Poland\label{AFFIL::PolandWUTElectronics}
\item Physics Program, Graduate School of Advanced Science and Engineering, Hiroshima University, 739-8526 Hiroshima, Japan\label{AFFIL::JapanUHiroshima}
\item School of Physics and Astronomy, Sun Yat-sen University, Zhuhai, China\label{AFFIL::ChinaUSunYatsen}
\item Department of Physical Sciences, Aoyama Gakuin University, Fuchinobe, Sagamihara, Kanagawa, 252-5258, Japan\label{AFFIL::JapanUAoyamaGakuin}
\item Division of Physics and Astronomy, Graduate School of Science, Kyoto University, Sakyo-ku, Kyoto, 606-8502, Japan\label{AFFIL::JapanUKyotoPhysicsandAstronomy}
\item Port d'Informaci\'o Cient{\'\i}fica, Edifici D, Carrer de l'Albareda, 08193 Bellaterrra (Cerdanyola del Vall\`es), Spain\label{AFFIL::SpainPIC}
\item INAF - Osservatorio Astrofisico di Torino, Strada Osservatorio 20, 10025  Pino Torinese (TO), Italy\label{AFFIL::ItalyOTorino}
\item Departamento de F{\'\i}sica, Universidad T\'ecnica Federico Santa Mar{\'\i}a, Avenida Espa\~na, 1680 Valpara{\'\i}so, Chile\label{AFFIL::ChileDepFisUTecnicaFedericoSantaMaria}
\item Faculty of Science, Ibaraki University, Mito, Ibaraki, 310-8512, Japan\label{AFFIL::JapanUIbaraki}
\end{enumerate}

\end{document}